\documentclass[sigconf,nonacm]{acmart}

\usepackage{multirow}
\usepackage{makecell}
\usepackage{amsmath}
\usepackage{mathrsfs}
\usepackage{bm}
\usepackage{amsfonts}
\usepackage{array}
\usepackage{algorithm}
\usepackage{algorithmic}
\usepackage{siunitx}
\usepackage{dcolumn}
\usepackage{float}
\usepackage{comment}
\usepackage{subfig}
\usepackage{graphicx}

\usepackage{tabularx}
\usepackage{booktabs}
\usepackage[table]{xcolor}
\usepackage{colortbl}
\usepackage{pifont}

\definecolor{groupgray}{RGB}{245,245,245}
\definecolor{ourblue}{RGB}{235,244,249}
\definecolor{bestpink}{RGB}{220,20,120}
\definecolor{secondblue}{RGB}{0,150,220}
\definecolor{ourblueA}{RGB}{232,242,250}
\definecolor{ourblueB}{RGB}{239,246,252}
\definecolor{ourblueC}{RGB}{246,250,254}

\newcommand{\cmark}{\ding{51}}
\newcommand{\xmark}{\ding{55}}

\AtBeginDocument{%
}

\acmYear{2026}
\acmConference[MM '26]
{the 34th ACM International Conference on Multimedia}
{November 10--14, 2026}
{Rio de Janeiro, Brazil}

\begin{document}

\title{SpeedyGS: Content-Aware 3D Gaussian Splatting Compression via Two-Stage Optimization}

\author{Junteng Zhang}
\affiliation{%
  \institution{Nanjing University}
  \city{Nanjing}
  \country{China}}
\email{zhangjunteng@smail.nju.edu.cn}

\author{Tong Chen}
\authornote{Corresponding author.}
\affiliation{%
  \institution{Nanjing University}
  \city{Nanjing}
  \country{China}}
\email{chentong@nju.edu.cn}

\author{Yuxin Zhao}
\affiliation{%
  \institution{Huawei Technologies}
  \city{Beijing}
  \country{China}}
\email{zhaoyuxin22@huawei.com}

\author{Yibo Shi}
\affiliation{%
  \institution{Huawei Technologies}
  \city{Beijing}
  \country{China}}
\email{shiyibo@huawei.com}

\author{Jing Wang}
\affiliation{%
  \institution{Huawei Technologies}
  \city{Beijing}
  \country{China}}
\email{wangjing215@huawei.com}

\author{Zhan Ma}
\affiliation{%
  \institution{Nanjing University}
  \city{Nanjing}
  \country{China}}
\email{mazhan@nju.edu.cn}

\renewcommand{\shortauthors}{J. Zhang et al.}

\begin{abstract}
  Recent progress in compressing large-scale 3D Gaussian Splatting (3DGS) data has substantially reduced storage footprint, network transmission bandwidth, and memory traffic to GPU caches before rendering. Yet decoding with advanced 3DGS codecs still takes seconds, making them unsuitable for interactive applications. To systematically address this challenge, we propose SpeedyGS, a Content-Aware 3DGS Compressor that separately optimizes the structural formation and statistical coding. First, in structural formation, we jointly optimize adaptive quantization and pruning under a unified rate–distortion objective, where the rate term is replaced by a lightweight rate proxy that estimates entropy coding cost of the next stage, thereby efficiently regulating Gaussian density and precision to yield a compact scene representation. Then, in the statistical coding phase, Gaussian geometry is converted into sparse octree tokens and subsequently undergoes multi-stage coding, while Gaussian attributes are serialized into a 1D token stream for entropy coding via a complexity‑controllable local autoregressive model. SpeedyGS achieves a favorable balance among optimization efficiency, compression performance, decoding latency, and rendering speed. Compared to vanilla 3DGS, SpeedyGS achieves up to 160$\times$ model size reduction with negligible quality degradation across common datasets. Compared to state-of-the-art compression methods, it also offers significantly faster decoding and accelerates optimization by 9$\times$ on consumer-grade hardware. To further reduce decoding overhead, the statistical coding stage also supports channel-wise, fixed-length coding for Gaussian as a simpler alternative, enabling SpeedyGS to better adapt to the underlying application and reduce decoding latency to nearly zero.
\end{abstract}

%%ACM M
\begin{CCSXML}
<ccs2012>
<concept>
<concept_id>10003752.10003809.10010031.10002975</concept_id>
<concept_desc>Theory of computation~Data compression</concept_desc>
<concept_significance>500</concept_significance>
</concept>
</ccs2012>
\end{CCSXML}

\ccsdesc[500]{Theory of computation~Data compression}

\keywords{3D Gaussian Splatting, Compression, Two-Stage Optimization}


\maketitle

\section{Introduction}
\label{sec:intro}

\begin{figure}[!t]
\centering 
\includegraphics[width=\linewidth]{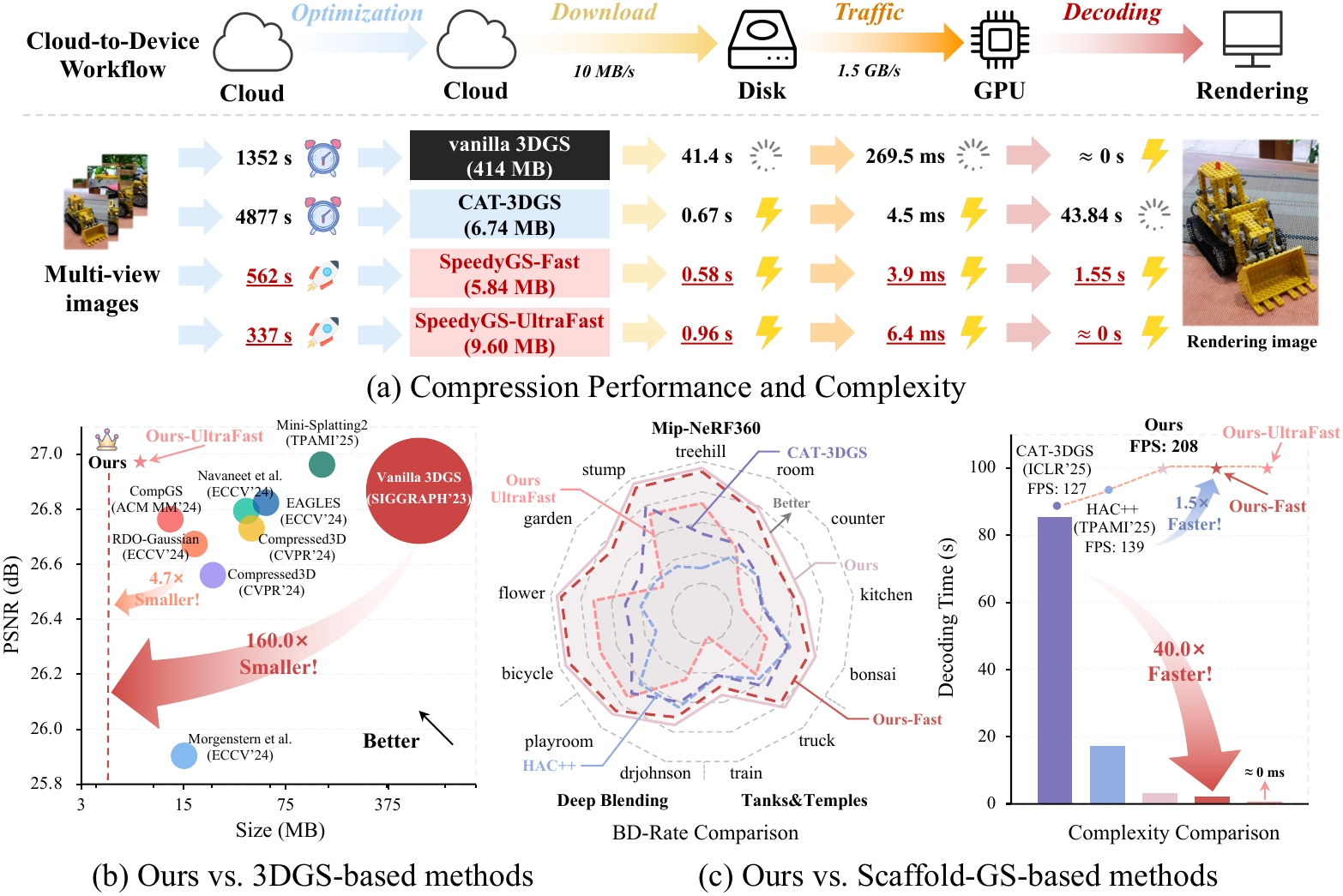}
\vspace{-6mm}
\caption{
(a) Cloud-to-device workflow: compressed 3DGS data must be decoded before rendering, making decoding speed as important as compression ratio. Relative to vanilla 3DGS~\cite{kerbl20233d}, SpeedyGS reduces optimization, storage, and download times. While rivaling the fast download speed of advanced codecs like CAT-3DGS~\cite{zhan2025cat3dgs}, it further delivers substantially lower decoding latency.
(b) Against 3DGS-based baselines, SpeedyGS attains the highest PSNR while maintaining the smallest storage. (c) Compared with Scaffold-GS-based methods, SpeedyGS offers improved compression efficiency (BD-Rate), reduced decoding complexity, and faster rendering. The Fast and UltraFast variants of SpeedyGS are obtained by simplifying the statistical coding stage.}
\vspace{-3mm}
\label{fig:first}
\end{figure}

Efficiently rendering complex 3D scenes with fine-grained details at interactive speed or faster remains a core challenge in computer graphics and computer vision. Recently, 3DGS~\cite{kerbl20233d} has gained prominence as a powerful technique for representing geometry and attributes by utilizing a dense collection of anisotropic Gaussians. This approach has been widely adopted across various applications, including digital human modeling, autonomous driving, and immersive telepresence~\cite{chen2024survey, wu2024recent}.

Nonetheless, the substantial data volume required for representing a scene with 3DGS presents a major obstacle to its swift market adoption. For example, the 3D Gaussian data generated for a typical scene in the Mip-NeRF360 dataset \cite{barron2022mip} surpasses 400 MB (see Fig.~\ref{fig:first}), incurring significant costs in terms of storage, network delivery, and memory traffic (\textit{e.g.}, fetching gigantic data from off-chip storage to on-chip cache for rendering or processing). Therefore, considerable efforts have been devoted to compressing 3DGS data. 

While some efforts have been made to design generalizable 3DGS coders \cite{chen2024fast,zhang2024compressing,huang2025hierarchical,yang2024benchmark,song2026tinysplat,zhang2026d}, the majority of endeavors has focused on per-scene optimization-based compression approaches due to their superior performance (and potentially faster decoding and rendering). These per-scene optimization methods essentially follow a ``training-as-compressing'' paradigm, and the method proposed in this paper also falls within this category.

\begin{figure}[t]
\centering 
\includegraphics[width=0.9\linewidth]{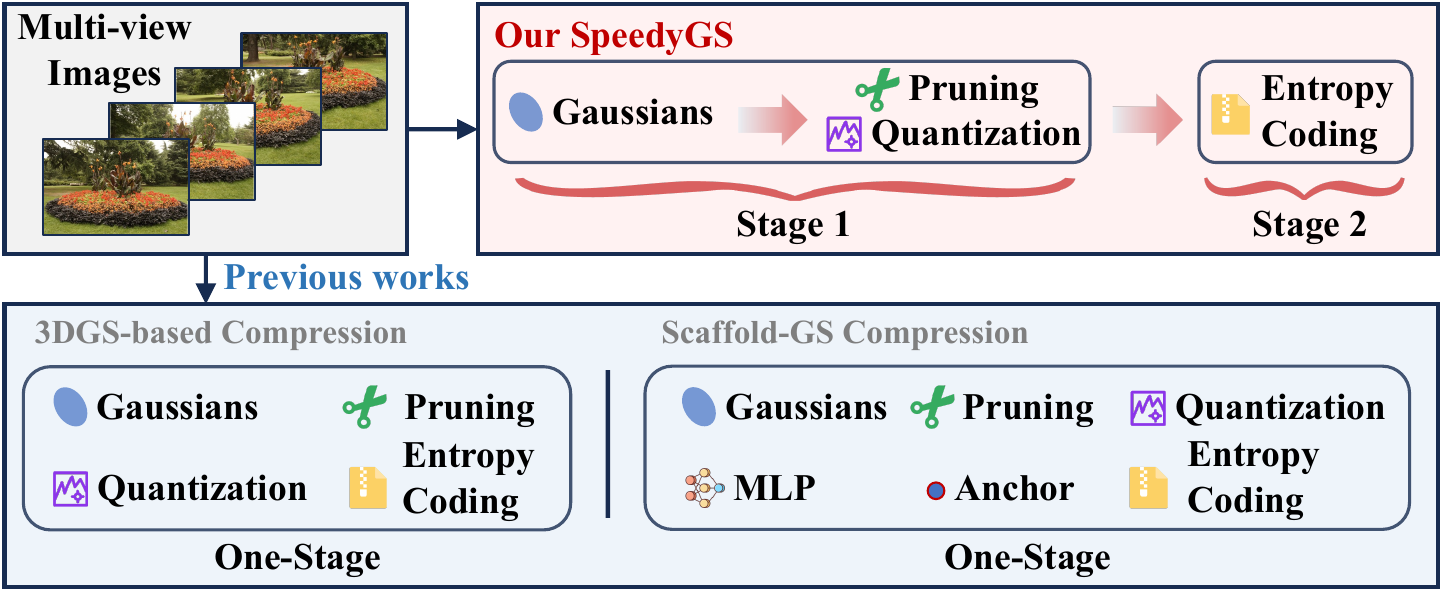}
\caption{Training pipeline for 3DGS compression. Unlike prior one-stage methods, we adopt a two-stage solution with pruning and quantization in Stage 1 and entropy coding in Stage 2.}
\vspace{-3mm}
\label{fig:difference}
\end{figure}

The top-performing solutions within the per-scene optimization category \cite{wang2024end,liu2024compgs,chen2024hac,chen2025hac++,wang2024contextgs,zhan2025cat3dgs,liu20253d,chen2026pcgs} typically adopt a one-stage approach that jointly optimizes structural formation\footnote{Structural formation is also referred to as ``compaction'' in \cite{bagdasarian20243dgs}.} (\textit{e.g.}, masked pruning, anchor-based prediction) and statistical coding (\textit{e.g.}, latent space feature quantization, entropy coding) under specific rate-distortion (R-D) criteria as in Fig.~\ref{fig:difference}. However, this one-stage optimization approach theoretically entails exploring a larger parameter space during training, which empirically leads to slower convergence (see Fig.~\ref{fig:vs_joint_optimization}). 
On the other hand, these methods tend to pursue high compression performance, while overlooking the decoding burden (which is unavoidable before rendering) and solely claiming ``rendering speed'' advantages. For example, the leading approach~\cite{zhan2025cat3dgs} requires $>$10 seconds to decode a compressed 3DGS scene with $\sim$500K anchor points before being piped for rendering, making it impractical for interactive applications.

\begin{figure*}[t] 
\centering 
\includegraphics[width=0.92\linewidth]{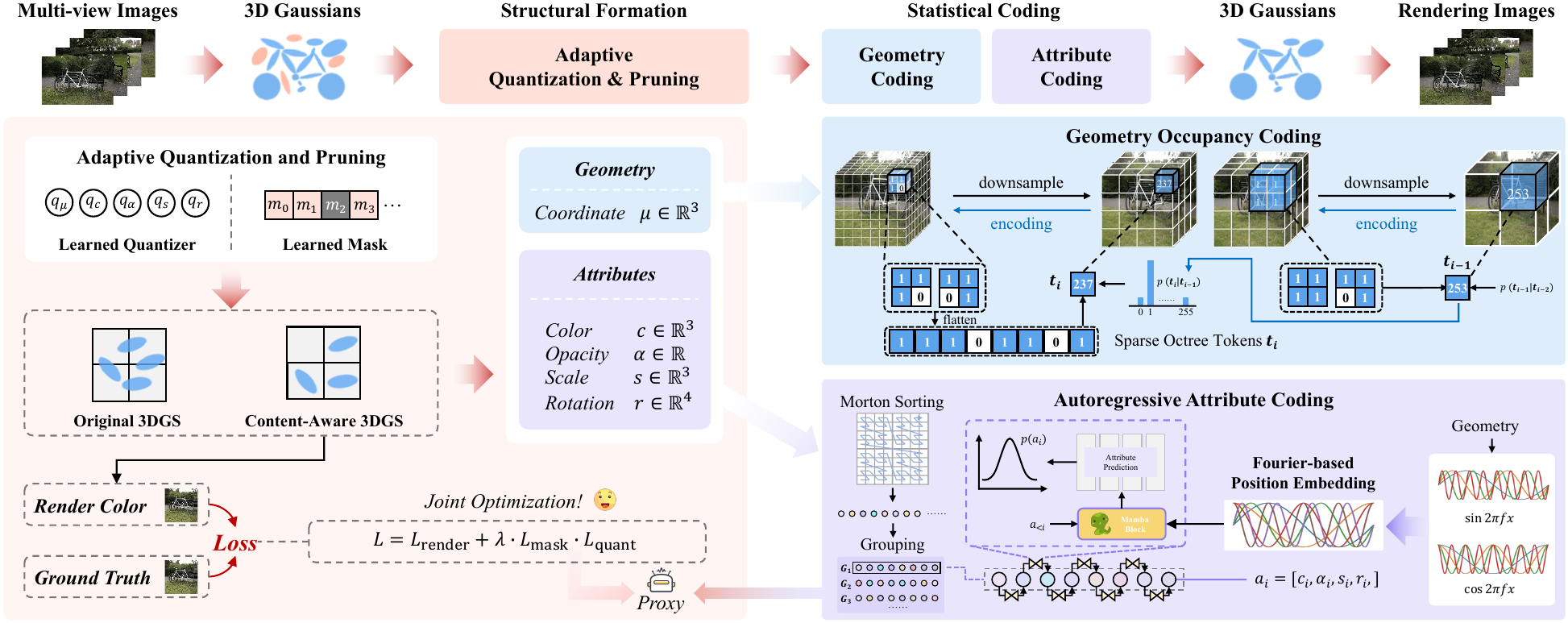} 
\vspace{-3mm}
\caption{SpeedyGS consists of two stages: structural formation and statistical coding. The former stage performs content-aware optimization of learnable quantizers and masks on Gaussian primitives, using a lightweight rate proxy of the entropy coding cost to produce a compact and coding-friendly representation (see Sec.~\ref{sec:struct_formation}). The latter stage applies statistical coding to further reduce redundancy. In the learned approach shown here (see Sec.~\ref{sec:learned_adaptive_codingtr}), Gaussian geometry is first encoded as sparse octree tokens, followed by a 1D sequence of Gaussian attributes conditioned on Fourier positional embeddings and local autoregressive neighbors. As an alternative, one can use channel-wise fixed-length codes instead (see Sec.~\ref{sec:fixed-length-codes}).}
\label{fig:pipeline} 
\end{figure*}

To address these challenges, we propose SpeedyGS, a content-aware 3D Gaussian Splatting compressor that decouples the original one-stage optimization into two sequential stages. As illustrated in Fig.~\ref{fig:pipeline}, this decomposition allows structural formation and statistical coding to be optimized separately, leading to a better and more flexible trade-off among optimization efficiency, compression performance, rendering speed, and decoding latency.

\begin{itemize}
    \item In the structural formation stage, our approach is to introduce content-aware adaptive quantization and pruning to jointly regulate Gaussian precision and density. This stage is optimized under R-D objective, where the rate term is approximated using a lightweight proxy designed to estimate the entropy coding cost of the subsequent stage, enabling formation of a compact and coding-friendly representation.
    \item In the statistical coding stage, we use learned adaptive coding: the Gaussian geometry (\textit{i.e.}, coordinates) is represented with sparse octree tokens and subsequently undergoes multi-stage coding, while Gaussian attributes such as color, opacity, scale, and rotation are linearized into a 1D sequence and encoded using parallelized local autoregression. This stage also supports channel-wise, fixed-length codes as an alternative to learned content-adaptive coding, enabling much faster decoding.
\end{itemize}

{SpeedyGS makes the following contributions:}
\begin{itemize}
    \item Motivated by the fundamental difference in the objectives of structural formation and statistical coding, we decouple the two stages, enabling stage-wise optimization with explicit content awareness. In particular, the formation stage proposes a lightweight rate proxy without actual entropy coding to optimize adaptive quantization and pruning for R-D performance, while the next statistical coding stage incorporates the complexity trade-off into the design of context-adaptive codes to reduce statistical redundancy.

    \item Upon the two-stage optimization, SpeedyGS also achieves leading performance. Compared to vanilla 3DGS at comparable perceptual quality, it achieves a compression ratio of approximately 160$\times$, a 2$\times$ improvement in optimization efficiency, rendering speed, and low decoding latency. It also offers a better overall trade-off than leading representative methods such as CAT-3DGS and HAC++. Additionally, SpeedyGS's two-stage optimization substantially reduces the training search space, achieving a notable 9$\times$ training speedup over CAT-3DGS as in Table~\ref{tab:quantitative_results}.

    \item SpeedyGS offers the flexibility to modify or replace individual stages as required. For instance, by parallelizing statistical attribute coding, SpeedyGS-Fast achieves a decoding latency of only 2.14 s (see Table~\ref{tab:quantitative_results}), while still outperforming CAT-3DGS and HAC++. SpeedyGS-UltraFast employs fixed-length codes, resulting in almost zero decoding latency while maintaining comparable compression performance to CAT-3DGS. 
\end{itemize}

\section{Related Work}
\label{sec:formatting}

\subsection{3DGS Structural Formation}
The vanilla 3DGS~\cite{kerbl20233d} exhibits an excessively dense spatial distribution of Gaussians, resulting in high storage and transmission costs as well as reduced rendering speed. To address this issue, structural formation techniques are commonly employed to refine the number and precision of Gaussians. Specifically, pruning methods aim to eliminate redundancy using strategies such as learned masking~\cite{lee2024compact, liu2025maskgaussian}, importance-based scoring~\cite{fan2024lightgaussian, girish2024eagles, fang2025efficient}, overlap analysis~\cite{papantonakis2024reducing}, and opacity ranking~\cite{zhang2025gaussianspa}, thereby reducing memory usage and accelerating rendering while maintaining fidelity.  Meanwhile, other methods reduce the precision of Gaussian properties through vector or scalar quantization~\cite{fan2024lightgaussian, lee2024compact, navaneet2024compgs, niedermayr2024compressed}. However, these studies often produce moderate compacting of 3DGS representations due to insufficient pruning, simplistic quantization strategies, and the independent handling of pruning and quantization.

Alternatively, Scaffold-GS~\cite{lu2024scaffold} introduced an anchor-based structure that clusters Gaussians and uses neural networks for predictive coding of associated primitives. However, it still retains a large number of Gaussians to preserve fidelity, hindering faster rendering.

\subsection{3DGS Statistical Coding}
Existing 3DGS compression techniques fall into two main categories: 3DGS-based and Scaffold-GS-based methods. 

3DGS-based methods reduce Gaussian redundancy using entropy coding and R-D optimization after structural formation of vanilla 3DGS~\cite{fan2024lightgaussian, niedermayr2024compressed,  wang2024end, liu2024compgs, chen2024fast}. Although faster rendering speed is attained, they mainly focus on the statistical properties of individual Gaussians, and only a small subset capture spatial correlations, thereby limiting compression performance.

Building on Scaffold-GS~\cite{lu2024scaffold}, several recent methods have achieved notable advancements by incorporating sophisticated context modeling techniques. These methods include capturing spatial dependencies through the fusion of hash grid features~\cite{chen2024hac}, leveraging context prediction based on local neighbors~\cite{wang2024contextgs}, utilizing channel-wise autoregressive models to exploit correlations within Gaussians~\cite{zhan2025cat3dgs, chen2025hac++}, \textit{etc}. Importantly, they typically perform one-stage optimization of structural formation and statistical coding, which either introduces additional hyperparameters~\cite{chen2024hac, zhan2025cat3dgs} or restricts pruning to local Gaussians~\cite{chen2025hac++}. 
Although such one-stage optimization achieves remarkable compression performance, it demands the exploration of a large parameter space during training, which greatly increases training effort. Additionally, employing complex context models in entropy coding often introduces a substantial decoding overhead.

In summary, existing 3DGS-based and Scaffold-GS-based compression methods mainly emphasize compression performance, while paying less attention to optimization and runtime complexity, which limits their suitability for practical applications. This calls for a new design that better balances optimization efficiency, compression performance, decoding latency, and rendering speed.

\section{Proposed Method}
\subsection{Preliminary}
% 3DGS
\textbf{3D Gaussian Splatting (3DGS)}~\cite{kerbl20233d} represents a 3D scene using a large set of anisotropic Gaussian primitives, each initialized using Structure‐from‐Motion (SfM) points. Every Gaussian is defined by its (geometry) coordinate $\boldsymbol\mu\in\mathbb R^3$ and a 3D covariance matrix $\boldsymbol\Sigma\in\mathbb R^{3\times3}$. 
The Gaussian at any spatial point $\boldsymbol x$ is thus:
\begin{equation}
    G(\boldsymbol x) \;=\;\exp\Bigl(-\tfrac12(\boldsymbol x-\boldsymbol\mu)^\top\boldsymbol\Sigma^{-1}(\boldsymbol x-\boldsymbol\mu)\Bigr).
\end{equation}

From a random viewpoint, corresponding 3D Gaussians are projected into a 2D image plane and rendered via a tile-based differentiable rasterization. Letting $N$ index the Gaussians sorted by depth at a given pixel position, the final color $\boldsymbol C$ is formed as
\begin{equation}
    \boldsymbol C \;=\;\sum\nolimits_{i \in N}\,\boldsymbol c_i\,\alpha_i\;\prod\nolimits_{j=1}^{i-1}\bigl(1-\alpha_j\bigr),
\end{equation}
where $\boldsymbol c_i$ and $\alpha_i$ are the view-dependent color of each Gaussian and its opacity, respectively. Our SpeedyGS is based on the above vanilla 3DGS representation.

\subsection{Problem Decomposition}
Early 3DGS compression resorted to pruning optimization and simple fixed-precision quantization, resulting in low efficiency. Later, numerous studies incorporated complex entropy modeling, leading to a one-stage optimization objective:
\begin{equation}
\min _{G, Q, M, \theta} [\underbrace{\mathcal{D}(G, Q, M)}_{\text {distortion}}+\lambda \underbrace{\mathcal{R}(G, Q, M ; \theta)}_{\text {rate}}], \label{eq:joint-opt}
\end{equation}
where $G$, $Q$, $M$, and $\theta$ denote the Gaussian parameters, quantization variables, masking variables, and coder parameters, respectively. In our implementation, for a scene with $N$ Gaussians, the one-stage optimization simultaneously searches over the $N\times59$ Gaussian parameters, $59$ quantization variables, $N\times2$ masking parameters, and $|\theta|$ coder parameters. Accordingly, the search space scales as
\begin{equation}
N\times59 + 59 +N\times2 +  |\theta|.
\end{equation}

One-stage optimization of $(G,Q,M)$ and $\theta$ leads to a substantially larger and more tightly coupled search space, since structural formation and statistical coding are entangled in a single objective. This increased coupling makes the optimization more difficult, often resulting in slower convergence and higher computational cost.

\begin{figure}[t]
\centering 
\includegraphics[width=0.9\linewidth]{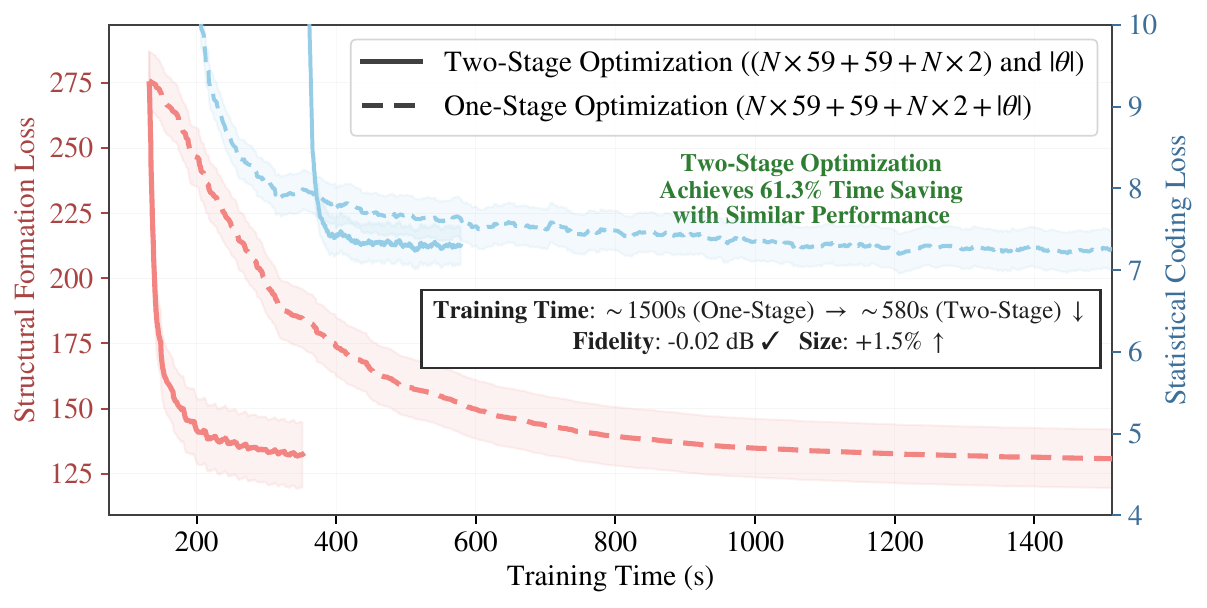}
\vspace{-3mm}
\caption{Training stability comparison between two-stage and one-stage optimization. The statistical coding loss corresponds to the final compressed bitstream size. Both methods are initialized from the same checkpoint at 10k iterations, after a shared densification--sparsification procedure. Solid lines denote the Structural Formation Loss, and dashed lines denote the Statistical Coding Loss. Results are shown on the \textit{bicycle} scene with $\lambda=2.5e-5$.}
\vspace{-5mm}
\label{fig:vs_joint_optimization}
\end{figure}

Recalling that optimizing $G$, $Q$, $M$, and $\theta$ serves different objectives: $G$, $Q$, and $M$ shape the 3DGS representation itself (via Gaussian parameter optimization, discretization, and sparsification), whereas $\theta$ optimizes the entropy model for statistical coding of that representation. Motivated by this distinction, we decompose the original one-stage optimization into two stages: structural formation and statistical coding, thereby reducing the search space of each stage to a more focused subproblem.

In structural formation, $(G, Q,M)$ are optimized first by:
\begin{equation}
\min _{G, Q, M} [\mathcal{D}_{\text {render }}(G, Q, M)+\lambda \mathcal{R}_{\text {formation }}(G, Q, M)], \label{eq:sf}
\end{equation}
where $\mathcal{D}_{\text {render }}$ is the rendering distortion between the rendered images and the ground truth. $\mathcal{R}_{\text {formation}}$ is the formation cost of Gaussians.

In statistical coding, we optimize $\theta$ with $G^*$, $Q^*$, and $M^*$ fixed by solving \eqref{eq:sf}:
\begin{equation}
\min _\theta \mathcal{R}_{\text {coding }}\left(G^*, Q^*, M^* ; \theta\right), \label{eq:sc}
\end{equation}
where $\mathcal{R}_{\text {coding }}$ is the rate of entropy coding. 

By replacing the monolithic one-stage optimization in \eqref{eq:joint-opt} with the two-stage decomposition in \eqref{eq:sf} and \eqref{eq:sc}, we substantially improve optimization efficiency. As shown in Fig.~\ref{fig:vs_joint_optimization}, the two-stage optimization achieves comparable reconstruction fidelity with much lower optimization cost, with only a negligible final gap of 0.02 dB, while reducing the optimization time by 61.3\%. The training curves also show faster convergence, indicating that the decomposition makes the optimization problem easier to solve by reducing the effective search space.

\subsection{Content-Aware Structural Formation} \label{sec:struct_formation}

As discussed earlier, we decouple structural formation from the statistical coding stage to reduce optimization complexity and accelerate convergence. However, this also prevents the formation stage from directly optimizing the true entropy coding cost. We therefore introduce a lightweight rate proxy to preserve rate-awareness during structural formation.

\begin{figure}[t]
\centering 
\subfloat
{\includegraphics[width=0.9\linewidth]{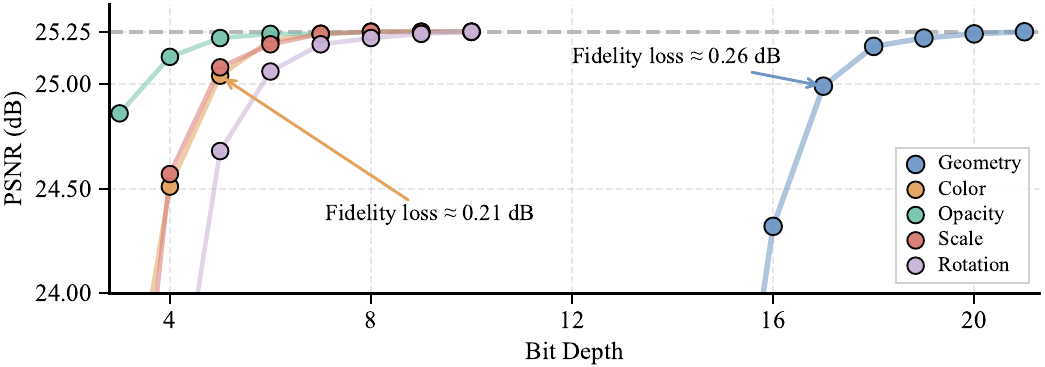}}\label{subfig:quantization}\\
\vspace{-3mm}
\subfloat
{\includegraphics[width=0.9\linewidth]{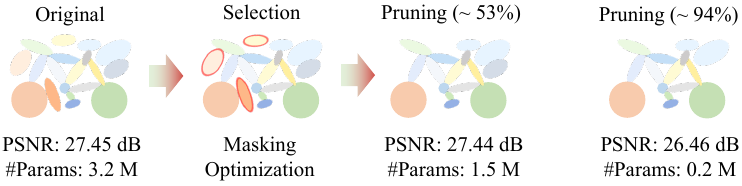}}\label{subfig:masking}\\
\vspace{-3mm}
\caption{Quantization and Pruning impact on fidelity.}
\vspace{-3mm}
\label{fig:quant}
\end{figure}

This proxy is realized through two complementary components: channel-wise quantization and Gaussian-wise pruning. Specifically, we introduce a learned channel-level quantizer $q_j$ for Gaussian primitives across the scene to enable dynamic precision adjustment for each channel, and a learned mask $\hat{M}_i$ for content-aware pruning. As shown in Fig.~\ref{fig:quant}, different Gaussian channels exhibit different sensitivity to quantization, while the fidelity also changes markedly under different pruning ratios, motivating adaptive quantization and adaptive pruning for compact 3DGS representation.

\subsubsection{Rate Proxy}

Existing learned compression methods~\cite{zhan2025cat3dgs, chen2024hac, chen2025hac++} typically define the rate term as the expected code length induced by an entropy model, which also corresponds to the statistical coding stage in our framework:
\begin{equation}
\mathcal{L}_{\mathrm{coding}}
=
\mathbb{E}\left[-\log_2 p_{\theta}(y \mid c)\right],
\end{equation}
where $p_{\theta}(y \mid c)$ is modeled by a conditional Laplace distribution.
Directly optimizing this objective during formation would require coupling structure optimization with context-dependent entropy coding, which substantially increases the optimization complexity.
We therefore seek a proxy of $\mathcal{L}_{\mathrm{coding}}$ for efficient optimization during structural formation.

As a first simplification, we remove the context dependency and model the centered variable $\tilde y = y - \mathrm{mean}(y)$ with a zero-mean Laplace distribution:
\begin{equation}
p(\tilde y; b)=\frac{1}{2b}\exp\left(-\frac{|\tilde y|}{b}\right),
\end{equation}
whose corresponding expected negative log-likelihood is
\begin{equation}
\label{eq:laplace}
\mathbb{E}\left[
\frac{|\tilde y|}{b \ln 2} + \log_2(2b)
\right],
\end{equation}
where $b$ is the scale parameter in the Laplace distribution.
This model avoids expensive context modeling while preserving the dependence of entropy coding cost on the signal scale. Under the assumed Laplace model, taking expectation with respect to $p(\tilde y;b)$ yields the differential entropy $h(\tilde y)$.

To further account for quantization, let $\hat y$ denote the discrete variable obtained by uniformly quantizing $\tilde y$ with step size $q$, and let $H(\hat y)$ denote its entropy. By the standard high-resolution quantization result~\cite{gray2002quantization}, with the derivation in the supplementary material,
\begin{equation}
H(\hat{y})
\approx
h(\tilde y)-\log_2 q
=
\log_2(2eb)-\log_2 q
\approx
\log_2 \frac{b}{q} + \mathrm{const},
\end{equation}
which indicates that, under the Laplace assumption, the entropy coding cost after quantization is mainly governed by the ratio between the signal scale $b$ and the quantization step size $q$.

To obtain a lightweight objective for the structural formation stage, we adopt the channel dynamic range as a coarse surrogate for the signal scale.
For the $j$-th channel of the Gaussian parameters $P_j$, including geometry and attributes, we define
\begin{equation}
\mathcal{L}_j
=
\log_2
\left\lceil
\frac{\max P_j - \min P_j}{q_j}
\right\rceil,
\end{equation}
where $q_j$ is the learnable quantization step size for that channel.
Summing over all channels gives the final quantization loss:
\begin{equation}
\mathcal{L}_{\mathrm{quant}}
=
\sum_j \mathcal{L}_j.
\end{equation}

In this way, $\mathcal{L}_{\mathrm{quant}}$ serves as our proposed quantization proxy, providing a lightweight channel-wise approximation of the entropy coding cost during structural formation.

Beyond per-channel quantization cost, the overall rate during structural formation also depends on the number of retained Gaussians. We therefore introduce a Gaussian-wise retention term into the rate proxy, so that it jointly accounts for both Gaussian precision and Gaussian count. To this end, each Gaussian $i$ is assigned a learnable score $s_i$, which is converted to a retention probability:
\begin{equation}
p_{i} = \sigma(s_{i})\,,
\end{equation}
where $\sigma(\cdot)$ is the sigmoid function. Based on $p_{i}$, we draw a soft, differentiable mask via Gumbel-Softmax:
\begin{equation}
M_{i} = \mathrm{GumbelSoftmax}\bigl(p_{i})\,,
\end{equation}
Then, the derived soft mask is binarized to a hard mask $\hat M_i \in {0,1}$, and gradients are back-propagated through $\hat{M}_i$ using masked rasterization~\cite{liu2025maskgaussian}. Let $\hat N$ denote the number of Gaussians before structural formation. We further define the mask loss as
\begin{equation}
\mathcal{L}_{\mathrm{mask}}
=
\frac{1}{\hat N}\sum_{i=1}^{N}\hat M_i,
\end{equation}
which measures the retention ratio of Gaussians. Using $\mathcal{L}_{\mathrm{mask}}$ as the count-related component, we define the overall rate proxy during structural formation as
\begin{equation}
\mathcal{R}_{\mathrm{proxy}}
=
\mathcal{L}_{\mathrm{mask}} \cdot \mathcal{L}_{\mathrm{quant}}.
\end{equation}
In this way, $\mathcal{R}_{\mathrm{proxy}}$ serves as a lightweight approximation of the entropy coding cost by jointly accounting for Gaussian count and precision. This enables content-aware optimization without introducing additional manually tuned hyperparameters~\cite{lee2024compact,chen2024hac,zhan2025cat3dgs}, automatically adapting the pruning strength to scene complexity.

\subsubsection{Structural Formation Loss}

Using the above rate proxy $\mathcal{R}_{\mathrm{proxy}}$ as the structural formation loss $\mathcal{L}_{\mathrm{formation}}$, the total training loss is:
\begin{equation}
\mathcal{L} = \mathcal{D}_{\mathrm{render}} + \lambda \cdot \mathcal{L}_{\mathrm{formation}},
\end{equation}
where $\mathcal{D}_{\mathrm{render}}$ denotes the rendering distortion between the rendered images and the ground truth.

\subsection{Complexity-Controllable Statistical Coding}

This stage further reduces statistical redundancy in the resulting Gaussians via entropy coding. This work presents two options: one utilizing fixed-length codes and the other employing learned codes.

\subsubsection{Fixed-length Coding} \label{sec:fixed-length-codes}
The resulting compact, well-organized representation from the structural formation stage supports channel-wise fixed-length coding: each channel’s quantized values are stored using a fixed bit width determined by its quantization precision. This design enables fully parallel decoding with almost zero latency. While metadata specifying each channel’s quantization precision must also be transmitted, the overhead is negligible, given that there are only 59 channels in total: 3 for geometry and 56 for attributes. SpeedyGS-UltraFast adopts this method.

 \subsubsection{Learned Adaptive Coding} \label{sec:learned_adaptive_codingtr}
We separate the encoding of Gaussian geometry and attributes, with $\mathcal{R}_{\mathrm{geo}}$ and $\mathcal{R}_{\mathrm{attr}}$ representing their rates, respectively. A low-complexity yet highly efficient framework, illustrated in Fig.~\ref{fig:pipeline}, is developed to fulfill this purpose. This framework is adopted in both SpeedyGS and SpeedyGS-Fast.

{\bf Geometry Occupancy Coding.} Fundamentally, due to its highly sparse density, Gaussians can be regarded as a LiDAR-like point cloud enriched with additional attribute dimensions, and thus can be compressed using point cloud compression (PCC) techniques. Existing work generally adopts the standard point cloud G-PCC~\cite{GPCC} for 3DGS geometry coding. While straightforward to implement, its performance is limited. 
Moreover, despite the high compression efficiency achieved by state-of-the-art AI-learned methods, they are often computationally intensive, requiring large models (\textit{e.g.}, 8M parameters~\cite{wang2025versatile}) and long inference times (\textit{e.g.}, more than 2 seconds per scene). Even the recent RENO~\cite{you2025reno}, which achieves 10 FPS processing speed at 12-bit LiDAR resolution, contains as many as 325K parameters, still posing considerable computational overhead for processing Gaussian geometry (typically between 18 and 20 bits after quantization).

Therefore, we specifically develop a per-scene optimized coding framework for Gaussian geometry. We convert geometry into compact, sparse octree tokens, as illustrated in Fig.~\ref{fig:pipeline}. Specifically, the Gaussian point cloud is first voxelized to a multiscale structure, with each unique point occupying a spatial position. In each downsampling step, high‑resolution voxels are aggregated into coarser voxels; the resulting high-resolution binary occupancy patterns are then mapped to 8-bit sparse octree tokens $\mathbf{t}_i$, ranging from 0 to 255. 
During both encoding and decoding, we feed the previous token $\mathbf{t}_{i-1}$ into a sparse CNN~\cite{choy20194d} with a softmax layer to model the conditional distribution $p(\mathbf{t}_{i}\mid \mathbf{t}_{i-1})$.

Overall, the geometry rate loss is estimated by:
\begin{equation}
\label{eq:loss_geo}
\mathcal{R}_{\mathrm{geo}}=-\sum_{i=1}^k\log p(\mathbf{t}_i\mid \mathbf{t}_{i-1}),
\end{equation}
where $p$ is the probability of sparse octree tokens $\mathbf{t}_i$, and $k$ is the maximum resolution of the Gaussian point cloud.

This sparse token-based process largely reduces computational complexity, resulting in a fast decoding speed. Meanwhile, benefiting from the per-scene training paradigm of 3DGS, we can specialize our model to each scene's unique geometry distribution, enabling effective learning with only 7.6K parameters, which is two orders of magnitude fewer than generalizable PCC coders~\cite{you2025reno}.

Note that voxelization in geometry coding may map multiple 3DGS points to the same position. In this case, only one point is explicitly encoded, while the number of duplicate points is recorded as metadata and compressed using LZMA~\cite{pavlov2009lzma}. As the number of duplicate points is small (about 1\% of the overall points), the resulting overhead is less than 1KB, which is negligible.

{\bf Autoregressive Attribute Coding.} Inspired by the success of image autoregressive models~\cite{van2016conditional} and recent transformer‐based tokenizers~\cite{dosovitskiyimage, yu2024image} that reshape high‐dimensional grids into 1D token sequences, we adopt a token-based paradigm for Gaussian attribute compression. Upon the decoded geometry positions, we use Morton ordering to sort the Gaussian functions, converting the 3D Gaussians into a 1D token sequence, as shown in Fig.~\ref{fig:pipeline}. Specifically, each token corresponds to one Gaussian and is defined by its full attribute vector:
\begin{equation}
\mathbf{a}_i =
\left[
\mathbf{c}_i,\;
\alpha_i,\;
\mathbf{s}_i,\;
\mathbf{r}_i
\right],
\end{equation}
where $\mathbf{c}_i$, $\alpha_i$, $\mathbf{s}_i$, and $\mathbf{r}_i$ denote the color, opacity, scale, and rotation attributes of the $i$-th Gaussian, respectively. In this way, the attributes of each Gaussian are modeled jointly as a single token, rather than as attribute channels. This spatially coherent ordering preserves locality, ensuring that neighboring Gaussians remain adjacent in the sequence and that local correlations among their attributes are retained, which facilitates effective autoregressive modeling with a causal State Space Model such as Mamba. To further improve the parallel decoding capability, we divide the 1D sequence into $G$ groups, each containing $W$ Gaussians.

{\it Fourier-based Positional Embedding}.
As analyzed, the geometry position is critical to our attribute coding. Thus, instead of using a simple linear embedding, we devise Fourier-based positional embeddings for Mamba SSM to map each geometry coordinate $\boldsymbol\mu$ across multiple frequency bands:
\begin{equation}
\phi(\boldsymbol\mu) = \big[\boldsymbol\mu;\, \sin(2\pi \mathbf{F} \boldsymbol\mu),\, \cos(2\pi \mathbf{F} \boldsymbol\mu) \big] \in \mathbb{R}^{3+2\times3L}
\end{equation}
with frequencies $\mathbf{F} = [f_0, f_1, \dots, f_{L-1}]$. $L$ denotes the number of frequency bands. This yields a high-dimensional vector that captures position at progressively finer scales. A learned linear projection then distills these rich cues down to $d$ dimensions:
\begin{equation}
\mathrm{PE}(\boldsymbol\mu)
= \mathrm{Linear}(\,\phi(\boldsymbol\mu))
\;\in\mathbb{R}^{d},
\end{equation}
In this work, we set $d=32$ after extensive experiments. 

As such, our autoregressive model can capture fine-grained spatial patterns while remaining compact.

\begin{table*}[t]
\centering
\caption{Quantitative comparison with representative methods on Mip-NeRF360. Scaffold-GS uses an anchor feature dimension of 50, consistent with HAC~\cite{chen2024hac}, CAT-3DGS~\cite{zhan2025cat3dgs}, and HAC++~\cite{chen2025hac++}. Two-Stage denotes two-stage optimization. For previous compression methods, complexity is reported at $\lambda_o=0.002$ following their original configurations, while ours is reported at the high-rate point ($\lambda=2.5e-5$). More methods and datasets are included in the supplementary material.}
\vspace{-3mm}
\label{tab:quantitative_results}
\setlength{\tabcolsep}{6pt}
\renewcommand{\arraystretch}{1.18}
{%
\resizebox{\linewidth}{!}{
\begin{tabular}{lcccc|cccc}
\toprule
& \multicolumn{4}{c|}{\textbf{Compression Performance}} & \multicolumn{4}{c}{\textbf{Complexity}} \\
\cmidrule(lr){2-5} \cmidrule(l){6-9}
\textbf{Method} & \textbf{SSIM} $\uparrow$ & \textbf{PSNR} $\uparrow$ & \textbf{LPIPS} $\downarrow$ & \textbf{Size (MB)} $\downarrow$ & \textbf{Two-Stage} & \textbf{Train (s)} $\downarrow$ & \textbf{Decode (s)} $\downarrow$ & \textbf{FPS} $\uparrow$ \\
\midrule

\multicolumn{9}{l}{\textit{\color{gray}\textbf{3DGS-based representation}}} \\
3DGS (SIGGRAPH'23)~\cite{kerbl20233d}
& 0.813 & 27.49 & 0.222 & 744.70 & -- & 1440 & -- & 124 \\
Mini-Splatting2 (TPAMI'25)~\cite{fang2025efficient}
& 0.821 & 27.43 & 0.214 & 159.18 & -- & 246 & -- & 201 \\
\midrule

\multicolumn{9}{l}{\textit{\color{gray}\textbf{Scaffold-GS-based representation and compression methods}}} \\
Scaffold-GS (CVPR'24)~\cite{lu2024scaffold}
& 0.812 & 27.81 & 0.228 & 227.95 & -- & 1088 & -- & 135 \\
HAC (ECCV'24)~\cite{chen2024hac}
& 0.807 & 27.53 & 0.238 & 15.26 & \xmark & 1803 & 11.82 & 133 \\
CAT-3DGS (ICLR'25)~\cite{zhan2025cat3dgs}
& 0.809 & 27.77 & 0.241 & 12.35 & \xmark & 5349 & 85.44 & 127 \\
HAC++ (TPAMI'25)~\cite{chen2025hac++}
& 0.807 &27.75 & 0.242 & 11.18 & \xmark & 2119 & 17.17 & 139 \\
\midrule

\multicolumn{9}{l}{\textit{\color{gray}\textbf{Our variants (3DGS-based)}}} \\
\rowcolor{ourblueA}
\textbf{Ours}
& 0.816 & 27.32 & 0.228 & \textbf{6.51} & \cmark & \textbf{559} & \textbf{3.27} & \textbf{208} \\
\rowcolor{ourblueB}
\textbf{Ours-Fast}
& 0.816 & 27.32 & 0.228 & 7.03 & \cmark & \textbf{538} & \textbf{2.14} & \textbf{208} \\
\rowcolor{ourblueC}
\textbf{Ours-UltraFast}
& 0.816 & 27.32 & 0.228 & 11.95 & \cmark & \textbf{313} & \textbf{$\approx 0$} & \textbf{208} \\
\bottomrule
\end{tabular}%
}
}
\end{table*}

\begin{figure*}[t]
  \centering
  \vspace{-4mm}
{\includegraphics[width=0.95\linewidth]{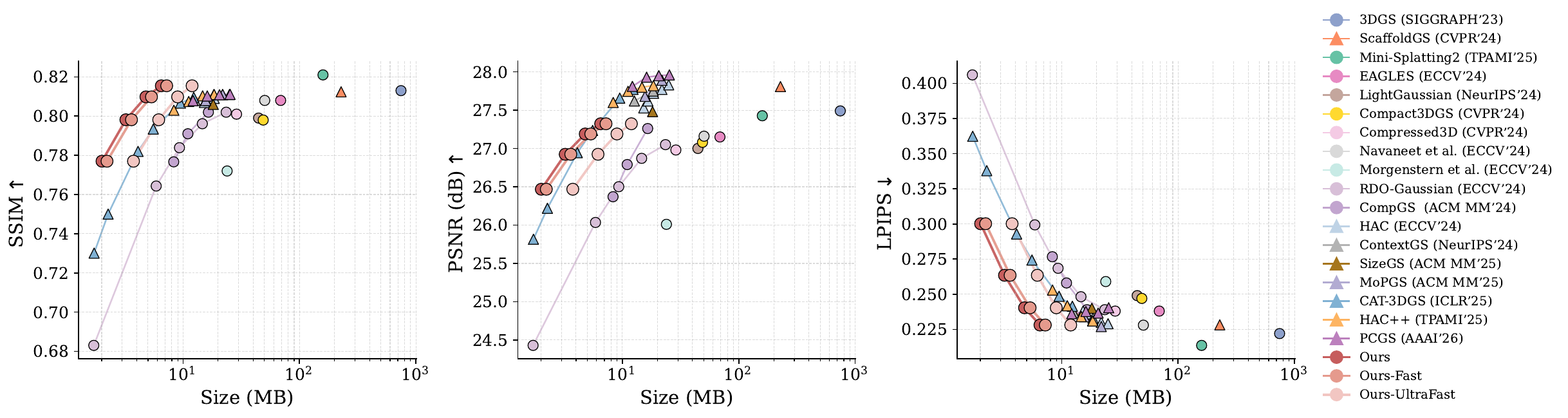}}
\vspace{-4mm}
  \caption{Rate-distortion curves on the Mip-NeRF360 dataset. Circular markers indicate 3DGS-based methods; triangular markers denote Scaffold-GS-based methods. Our method builds on Mini-Splatting2~\cite{fang2025efficient} (green circular).} 
  \vspace{-1mm}
  \label{fig:rd_curve_main}
\end{figure*}

{\it Autoregressive Prediction Coding.}
The Gaussian attribute sequence is then fed into a causal Mamba SSM architecture~\cite{dao2024transformers}, producing an autoregressive coding model:
\begin{equation}
p(\boldsymbol{a}_{1:W})
=\prod_{i=1}^W
p\bigl(\boldsymbol{a}_i \mid \mathrm{Mamba}(\boldsymbol{a}_{<i},\,\mathrm{PE}(\boldsymbol\mu_{\le i}))\bigr),
\end{equation}
where $\boldsymbol{a}$ is the attribute of each Gaussian, and $W$ is the number of Gaussians in each group. The compression rate is:
\begin{equation}
\label{eq:loss_attr}
\mathcal{R}_{\mathrm{attr}} = -\frac{1}{G}\sum\nolimits_{i=1}^G \log_2
p\bigl(\boldsymbol{a}_i),
\end{equation}
where $G$ is the total number of groups.

\section{Experiments \& Results}

\subsection{Experimental Setup}
\subsubsection{Implementation}
We implemented SpeedyGS in PyTorch using vanilla 3DGS~\cite{kerbl20233d} as the basic representation, built upon Mini-Splatting2~\cite{fang2025efficient} (an improved version of vanilla 3DGS), and evaluated it, along with other baselines, on a single consumer-grade GPU. We set the regularization weight $\lambda$ with various values from \{2e-4, 1e-4, 5e-5, 2.5e-5\} for rate-distortion trade‑offs. Training uses 30,000 iterations, including 5,000 for statistical coding. To save memory, we follow HAC~\cite{chen2024hac} and randomly sample 5\% of Gaussians during statistical coding. For efficiency, we adopt the optimizer scheduling strategy in~\cite{mallick2024taming, ren2025fastgs}, which includes separated spherical harmonics (SH) rendering and an efficient Adam update scheme. More details are provided in the supplementary material.

We set the number of frequency bands $L=12$ and the number of Gaussians in each group of the attribute coding $W=256$ to obtain SpeedyGS's main model. To accelerate the decoding speed, we reduce the number of Gaussians in each group to $W=32$, which enhances model parallelism and yields our Fast version. In addition, our UltraFast version uses only channel-wise fixed‑length codes (\textit{i.e.}, ultra‑fast entropy coding) instead of context modeling.

\subsubsection{Datasets \& Metrics} 
To thoroughly assess performance, we follow established conventions to evaluate the models on three large-scale datasets: Mip-NeRF360~\cite{barron2022mip}, Tanks\&Temples~\cite{knapitsch2017tanks}, and DeepBlending~\cite{hedman2018deep}.  We also report R-D performance (\textit{e.g.}, plotting R-D curves and calculating BD-Rate~\cite{bjontegaard2001calculation}).

\subsubsection{Benchmarking Baselines}
Comparative baselines include vanilla 3DGS~\cite{kerbl20233d}, Scaffold-GS~\cite{lu2024scaffold}, nine 3DGS-based approaches~\cite{fang2025efficient,girish2024eagles,fan2024lightgaussian,lee2024compact,niedermayr2024compressed,navaneet2024compgs,morgenstern2024compact,wang2024end,liu2024compgs}, and seven Scaffold-GS-based approaches~\cite{chen2024hac,chen2025hac++,chen2026pcgs,wang2024contextgs,zhan2025cat3dgs,liu20253d,xie2025sizegs}. Among them, methods without joint R-D optimization are shown as single R-D points. Note that the Scaffold-GS anchor feature dimension is set to 50, to align with prior compression baselines such as HAC++~\cite{chen2025hac++} and CAT-3DGS~\cite{zhan2025cat3dgs}.

\subsection{Comparative Studies}
\subsubsection{Quantitative Evaluation}
Figure~\ref{fig:rd_curve_main} and Table~\ref{tab:quantitative_results} demonstrate the advanced performance of SpeedyGS on the Mip-NeRF360. Compared with vanilla 3DGS, our full model reduces the scene size from 744.70\,MB to 6.51\,MB, corresponding to a compression ratio of about 114$\times$, while also increasing rendering speed from 124 to 208 FPS. Our lightweight variants still achieve high compression ratios, with Ours-Fast and Ours-UltraFast reaching about 106$\times$ and 62$\times$, respectively. Similar trends are observed on Tanks\&Temples and DeepBlending, as reported in the supplementary material.

Compared with previous compression methods, SpeedyGS achieves the smallest storage size among all compared methods in Table~\ref{tab:quantitative_results}, demonstrating superior compression efficiency. In particular, on Mip-NeRF360, SpeedyGS achieves a 50.52\% BD-Rate gain over CAT-3DGS and a 71.14\% BD-Rate gain over HAC on the SSIM metric. 

Beyond compression, SpeedyGS also offers clear complexity advantages. Compared with CAT-3DGS, it reduces the training time from 5349 s to 559 s, which further reflects that the proposed two-stage optimization helps reduce the optimization search space and improve training efficiency. Compared with HAC++, it further lowers the decoding time from 17.17\,s to 3.27\,s. Meanwhile, SpeedyGS achieves the highest rendering speed of 208 FPS among all compared methods. Note that HAC, CAT-3DGS, and HAC++ are Scaffold-GS-based methods, and the conversion from Scaffold-GS back to the vanilla 3DGS representation introduces additional delay (\textit{e.g.}, about 5\,ms on our testing platform) before rendering.

\subsubsection{Qualitative Evaluation}
\begin{figure*}[t] 
\centering 
\includegraphics[width=0.92\linewidth]{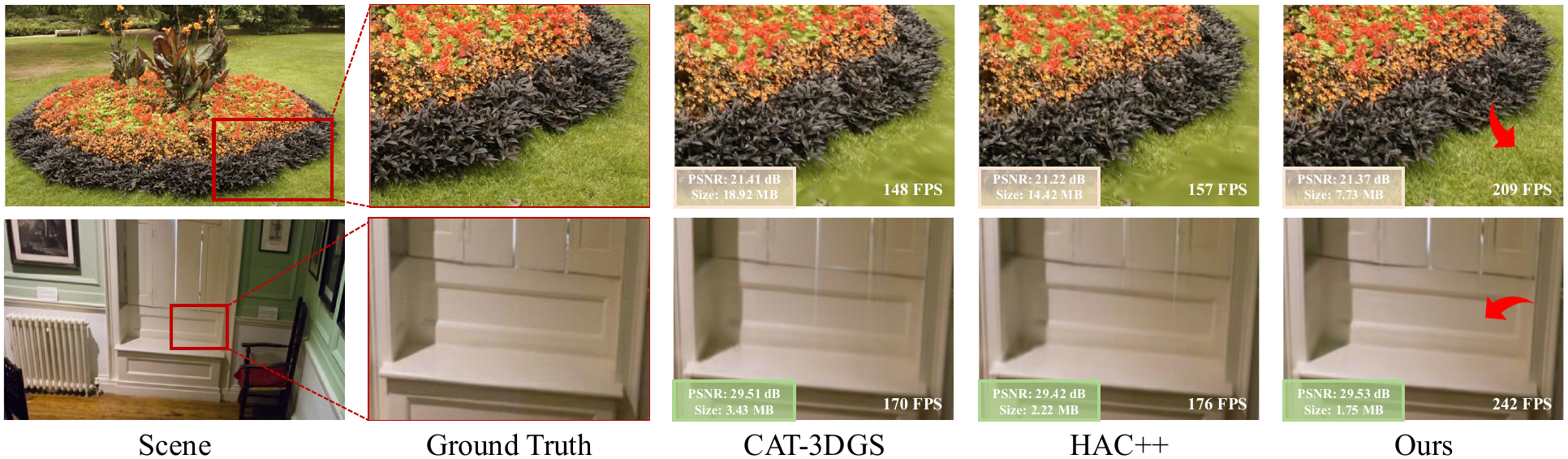} 
\vspace{-2mm}
\caption{Visualization comparison between SpeedyGS and baselines, including vanilla 3DGS~\cite{kerbl20233d}, CAT-3DGS~\cite{zhan2025cat3dgs}, and HAC++~\cite{chen2025hac++}, on \textit{flower} from the Mip-NeRF360 dataset and \textit{drjohnson} from DeepBlending dataset. Please zoom in to examine the details.}
\vspace{-2mm}
\label{fig:visual} 
\end{figure*}

Figure~\ref{fig:visual} presents qualitative comparisons with CAT-3DGS and HAC++. SpeedyGS achieves higher compactness while delivering better visual quality. For example, in the ``flower'' scene, prior methods fail to render the grass, whereas SpeedyGS reconstructs it clearly. Additional qualitative results are provided in the supplementary material.

\subsection{Ablation Study}

\begin{table}[t]
\centering
\caption{Ablation study of the proposed design on Mip-NeRF360. The anchor denotes the full SpeedyGS model. The BD-Rate reported here is computed based on PSNR.}
\vspace{-3mm}
\label{tab:ablation}
\setlength{\tabcolsep}{5.5pt}
\renewcommand{\arraystretch}{1.18}
\resizebox{\linewidth}{!}{%
\begin{tabular}{lcccc}
\toprule
\textbf{Ablation Item} & \textbf{BD-Rate} $\downarrow$ & \textbf{Train (s)} $\downarrow$ & \textbf{Decode (s)} $\downarrow$ & \textbf{FPS} $\uparrow$ \\
\midrule
\rowcolor{ourblueA}
\textbf{SpeedyGS (anchor)} & \textbf{0.00\%} & \textbf{559} & \textbf{3.27} & \textbf{208} \\\hline
\multicolumn{5}{l}{\textit{\color{gray}\textbf{Optimization strategy}}} \\
w/o two-stage optimization & -1.81\% & 1429 & 3.28 & 206 \\
\midrule

\multicolumn{5}{l}{\textit{\color{gray}\textbf{Rate proxy}}} \\
w/o masking & +79.03\% & 618 & 2.92 & 181 \\
w/ additive masking $\alpha=0.0001$ & +18.44\% & 583 & 3.36 & 193 \\
w/ additive masking $\alpha=0.0005$ & +4.94\% & 563 & 3.34 & 206 \\
w/ additive masking $\alpha=0.001$  & +12.64\% & 541 & 3.48 & 214 \\
w/ additive masking $\alpha=0.002$  & +5.03\% & 531 & 3.53 & 218 \\
\midrule

\multicolumn{5}{l}{\textit{\color{gray}\textbf{Statistical coding}}} \\
w/o statistical coding & +89.08\% & 313 & $\approx 0$ & 208 \\
w/o attribute coding & +67.60\% & 432 & 0.24 & 208 \\
w/o geometry coding & +28.10\% & 440 & 3.03 & 208 \\
w/ G-PCC & +0.12\% & 440 & 10.06 & 208 \\
\midrule

\multicolumn{5}{l}{\textit{\color{gray}\textbf{Training acceleration}}} \\
w/o optimizer schedule & +1.10\% & 1143 & 3.24 & 209 \\
% \midrule

\bottomrule
\end{tabular}%
}
\vspace{-4mm}
\end{table}

{\bf Optimization Strategies.}
Table~\ref{tab:ablation} validates the effectiveness of our two-stage optimization. Compared with our two-stage design, directly optimizing structural formation and statistical coding together achieves a slightly better BD-rate (-1.81\%), but increases the training time from 559\,s to 1429\,s, which is 2.56$\times$ slower. This suggests that one-stage optimization can bring a marginal coding gain, but at the cost of a substantially larger search space.

\begin{figure}[t]
\centering
\subfloat[]{\includegraphics[width=0.48\linewidth]{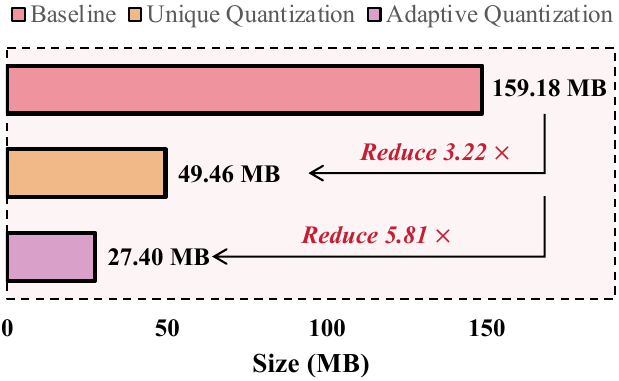}\label{subfig:ablation_quant}}
\hspace{2mm}
\subfloat[]{\includegraphics[width=0.48\linewidth]{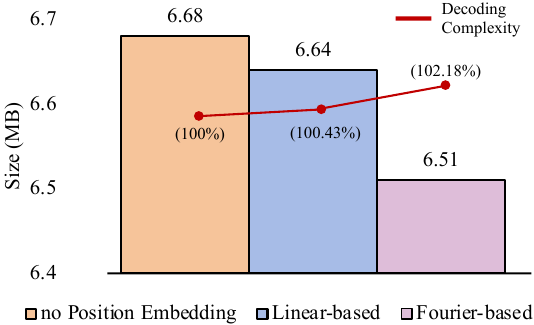}\label{subfig:ablation_PE}}
\vspace{-3mm}
\caption{Ablation studies on the Mip-NeRF360 dataset: (a) adaptive quantization, and (b) positional embedding.}
\vspace{-5mm}
\label{fig:ablation_study}
\end{figure}

{\bf Rate Proxy.} Table~\ref{tab:ablation} and Fig.~\ref{subfig:ablation_quant} evaluate the impact of the proposed rate proxy from two aspects: the masking strategy and the quantization scheme. For masking, we compare no masking, additive masking as an additional loss~\cite{lee2024compact,chen2024hac,zhan2025cat3dgs}, and our content-aware multiplicative masking. Additive masking improves compression performance and rendering speed, but its behavior is unstable and depends heavily on the manually tuned hyperparameter $\alpha$. In contrast, our multiplicative formulation jointly couples masking and quantization in a single rate proxy, eliminating manual hyperparameter tuning while consistently delivering strong performance.

For quantization, we compare uniform and adaptive schemes in Fig.~\ref{subfig:ablation_quant}. Using uniform quantization, where the precisions of geometry and attributes are fixed to 20-bit and 10-bit, respectively, results in a compressed size of 49.46 MB. Replacing it with the proposed adaptive quantization reduces the size to 27.40 MB, demonstrating the effectiveness of the quantization component in the rate proxy.

In addition, we also compare different rate proxy designs as shown in Eq.~(\ref{eq:laplace}), and find that the proposed quantization proxy better balances compression performance and training efficiency. Detailed comparisons are provided in the supplementary material.

{\bf Statistical Coding.}
Table~\ref{tab:ablation} evaluates the effectiveness of our statistical coding framework. Here, ``w/o Statistical Coding'' means that we do not use the proposed learned statistical coding modules, and instead apply fixed-length codes for ultra-fast coding. Although this setting yields nearly zero decoding latency, the BD-rate degrades significantly by 89.08\%, indicating that fixed-length coding alone cannot sufficiently remove the statistical redundancy remaining after structural formation.

For geometry coding, replacing our learned geometry coder with G-PCC causes slight performance degradation and substantially higher decoding complexity. Geometry decoding takes 0.24 s with our method versus 7.03 s with G-PCC, making ours about 29$\times$ faster with nearly the same compression performance. For attribute coding, removing the learned attribute coder and keeping only fixed-length coding causes a severe BD-rate degradation of 67.60\%, indicating that attribute coding contributes most of the coding gain.

{\bf Training Acceleration.}
For efficiency, we adopt the optimizer scheduling strategy of~\cite{mallick2024taming,ren2025fastgs}, including separated spherical harmonics (SH) rendering and an efficient Adam update scheme. It reduces the implementation overhead of Gaussian optimization, especially the frequent updates of higher-order SH coefficients. Removing it increases training time from 559 s to 1143 s.

{\bf Positional Embedding.}
Figure~\ref{subfig:ablation_PE} compares linear positional embedding with Fourier‑based in the autoregressive attribute coder. It can be seen that our embedding improves performance (reducing size from 6.68 to 6.51 MB) with negligible complexity.

\section{Conclusion}
This paper introduces SpeedyGS, a two-stage, content-aware 3DGS compression framework. In the first stage, SpeedyGS performs content-aware structural formation using adaptive quantization and pruning under a proxy-based rate-distortion objective. The second stage then applies complexity-controllable statistical coding to further reduce statistical redundancy. One can choose learned adaptive coding, which consists of sparse octree token–based geometry compression and 1D token sequence–based attribute compression, or opt for channel-wise, fixed-length coding. By decoupling the pipeline, this design avoids the large parameter space of monolithic one-stage optimization and enables efficient, independent optimization within each stage. Experiments demonstrate that SpeedyGS achieves a favorable balance among optimization efficiency, compression performance, decoding latency, and rendering speed compared to prior methods.

Currently, SpeedyGS's UltraFast version already supports real-time decoding. Our future work will focus on accelerating our full and fast versions for real-time processing.

\bibliographystyle{ACM-Reference-Format}
\bibliography{sample-base}

\clearpage
\appendix
\definecolor{groupgray}{RGB}{245,245,245}
\definecolor{ourblue}{RGB}{235,244,249}
\definecolor{bestpink}{RGB}{220,20,120}
\definecolor{secondblue}{RGB}{0,150,220}
\definecolor{ourblueA}{RGB}{232,242,250}
\definecolor{ourblueB}{RGB}{239,246,252}
\definecolor{ourblueC}{RGB}{246,250,254}

% \AtBeginDocument{%
%   \providecommand\BibTeX{{%
%     Bib\TeX}}}
    
% \settopmatter{printacmref=false}
% \setcopyright{none}
% \renewcommand\footnotetextcopyrightpermission[1]{}
% \pagestyle{plain}

% \acmConference[MM '26]{the 34th ACM International Conference on Multimedia}{November 10--14, 2026}{Rio de Janeiro, Brazil}

\title{Supplementary Materials of ``SpeedyGS: Content-Aware 3D Gaussian Splatting Compression via Two-Stage Optimization''}

% \maketitle

\section{From Coding Loss to Quantization Proxy}
\label{sec:appendix_quant_proxy}

In this section, we explain how the quantization loss used in the structural formation stage is motivated from the statistical coding objective in our two-stage framework. As discussed in the main paper, decoupling structural formation from statistical coding improves optimization efficiency, but it also prevents the former stage from directly optimizing the true entropy coding cost of the latter stage. We therefore introduce a lightweight quantization proxy that preserves rate awareness during structural formation, while remaining much cheaper to optimize than full context-dependent entropy coding.

\subsection{Coding Loss Under an Entropy Model}

In learned compression, the rate term is usually defined as the expected code length induced by an entropy model. For a symbol $y$ with context $c$, the coding cost can be written as
\begin{equation}
\mathcal{L}_{\mathrm{coding}}
=
\mathbb{E}\left[-\log_2 p_\theta(y \mid c)\right],
\label{eq:app_coding_loss}
\end{equation}
where $p_\theta(y\mid c)$ denotes the conditional probability assigned by the entropy model.

In our framework, this is the rate term optimized in the statistical coding stage. However, directly using Eq.~\eqref{eq:app_coding_loss} during structural formation would re-couple structure optimization with context-dependent entropy modeling, which would substantially increase optimization complexity. We therefore seek a lightweight proxy that captures the dominant trend of coding cost while being more suitable for the structural formation stage.

\subsection{A Zero-Mean Laplace Proxy}

As a first simplification, we remove the context dependency in Eq.~\eqref{eq:app_coding_loss}. Instead of modeling the conditional distribution $p_\theta(y\mid c)$, we consider a simpler unconditional model for the centered signal
\begin{equation}
\tilde y = y - \mathrm{mean}(y).
\end{equation}
We model $\tilde y$ using a zero-mean Laplace distribution:
\begin{equation}
p(\tilde y;b)=\frac{1}{2b}\exp\left(-\frac{|\tilde y|}{b}\right),
\label{eq:app_laplace_pdf}
\end{equation}
where $b>0$ is the scale parameter.

Under this model, the coding cost of one value is approximated by its negative log-likelihood in bits:
\begin{align}
-\log_2 p(\tilde y;b)
&=
-\log_2\left(\frac{1}{2b}\exp\left(-\frac{|\tilde y|}{b}\right)\right) \\
&=
\log_2(2b)+\frac{|\tilde y|}{b\ln 2}.
\label{eq:app_laplace_nll}
\end{align}
Averaging over all centered values gives the Laplace proxy
\begin{equation}
\mathcal{L}_{\mathrm{Lap}}
=
\mathbb{E}\left[
\frac{|\tilde y|}{b\ln 2}+\log_2(2b)
\right].
\label{eq:app_lap_proxy}
\end{equation}

Compared with the original coding loss, $\mathcal{L}_{\mathrm{Lap}}$ avoids expensive context modeling while preserving the dependence of coding cost on signal scale. The next step is to relate this scale-dependent cost to quantization.

\subsection{Why Quantization Step Size Matters}

After quantization, the coding cost depends not only on signal variation, but also on the quantization step size. Consider a continuous variable $\tilde y$ that is uniformly quantized with step size $q$, producing a discrete variable $\hat y$. A standard high-resolution quantization result~\cite{gray2002quantization} states that
\begin{equation}
H(\hat y)\approx h(\tilde y)-\log_2 q,
\label{eq:app_hrq}
\end{equation}
where $H(\hat y)$ is the entropy after quantization and $h(\tilde y)$ is the differential entropy of the original continuous variable.

For the zero-mean Laplace model in Eq.~\eqref{eq:app_laplace_pdf}, the differential entropy is
\begin{equation}
h(\tilde y)=\log_2(2eb).
\label{eq:app_laplace_entropy}
\end{equation}
Substituting Eq.~\eqref{eq:app_laplace_entropy} into Eq.~\eqref{eq:app_hrq}, we obtain
\begin{align}
H(\hat y)
&\approx
\log_2(2eb)-\log_2 q \\
&=
\log_2 \frac{b}{q} + \log_2(2e).
\label{eq:app_entropy_bq}
\end{align}
Therefore,
\begin{equation}
H(\hat y)\approx \log_2 \frac{b}{q}+\mathrm{const}.
\label{eq:app_entropy_bq_const}
\end{equation}

This shows that, under the Laplace assumption, the coding cost after quantization is mainly governed by the ratio between signal scale $b$ and quantization step size $q$.

\subsection{From a Scale-Based View to a Range-Based Proxy}

Equation~\eqref{eq:app_entropy_bq_const} indicates that the coding cost after quantization is largely determined by the ratio between signal scale and quantization step size. However, explicitly estimating a Laplace scale parameter for every channel during structural formation is unnecessary. Following the design principle of the main paper, we instead adopt a lightweight channel-wise approximation.

For the $j$-th channel of Gaussian parameters $P_j$, we use its dynamic range
\begin{equation}
R_j = \max P_j - \min P_j
\label{eq:app_range_def}
\end{equation}
as a coarse surrogate for signal variation. Although $R_j$ is not identical to the Laplace scale parameter $b$, both reflect the overall spread of the signal. Motivated by this relation, we replace the scale-dependent term in Eq.~\eqref{eq:app_entropy_bq_const} with the range-based surrogate
\begin{equation}
\log_2 \frac{b}{q_j}
\quad \Longrightarrow \quad
\log_2 \frac{R_j}{q_j}.
\label{eq:app_b_to_r}
\end{equation}

This surrogate also admits a simple interpretation: the ratio $R_j/q_j$ reflects how many quantization intervals are spanned by the $j$-th channel, and thus provides a coarse measure of its quantization complexity. Accordingly, we define the channel-wise quantization proxy as
\begin{equation}
\mathcal{L}_j
=
\log_2
\left\lceil
\frac{\max P_j - \min P_j}{q_j}
\right\rceil,
\end{equation}
where $q_j$ is the learnable quantization step size for channel $j$. Summing over all channels gives
\begin{equation}
\mathcal{L}_{\mathrm{quant}}
=
\sum_j \mathcal{L}_j.
\label{eq:app_quant_proxy}
\end{equation}

Therefore, $\mathcal{L}_{\mathrm{quant}}$ serves as a lightweight quantization proxy during structural formation. It is simple to compute, encourages compact and quantization-friendly representations, and aligns with the rate-aware design of the main paper without introducing full entropy coding into this stage.

\subsection{Comparison with an Alternative Rate Proxy}

To further validate the proxy design, we compare the proposed range-based quantization proxy in Eq.~\eqref{eq:app_quant_proxy} with the scale-based zero-mean Laplace proxy in Eq.~\eqref{eq:app_lap_proxy} under the same decomposed optimization framework. Specifically, we replace the proposed proxy in the structural formation stage with the Laplace proxy while keeping all other components unchanged. As shown in Table~\ref{tab:ablation_proxy}, the Laplace variant achieves a slightly BD-Rate gain, with nearly unchanged decoding speed and rendering FPS. Nevertheless, we retain the proposed range-based proxy in the final design because it enables faster training and is better aligned with the goal of efficient structural formation.

\begin{table}[t]
\centering
\caption{Comparison between the proposed range-based quantization proxy in Eq.~\eqref{eq:app_quant_proxy} and the scale-based Laplace proxy in Eq.~\eqref{eq:app_lap_proxy} on Mip-NeRF360 under the same decomposed optimization framework. The BD-Rate is computed based on PSNR.}
\vspace{-3mm}
\label{tab:ablation_proxy}
\setlength{\tabcolsep}{5.5pt}
\renewcommand{\arraystretch}{1.18}
\resizebox{\linewidth}{!}{%
\begin{tabular}{lcccc}
\toprule
\textbf{Rate Proxy} & \textbf{BD-Rate} $\downarrow$ & \textbf{Train (s)} $\downarrow$ & \textbf{Decode (s)} $\downarrow$ & \textbf{FPS} $\uparrow$ \\
\midrule
\rowcolor{ourblueA}
\textbf{Range-based proxy} & \textbf{0.00\%} & \textbf{559} & \textbf{3.27} & \textbf{208} \\
Scale-based Laplace proxy & -0.43\% & 663 & 3.25 & 205 \\
\bottomrule
\end{tabular}%
}
\vspace{-4mm}
\end{table}

\section{Datasets}
\begin{figure}[ht] 
\centering 
\includegraphics[width=\linewidth]{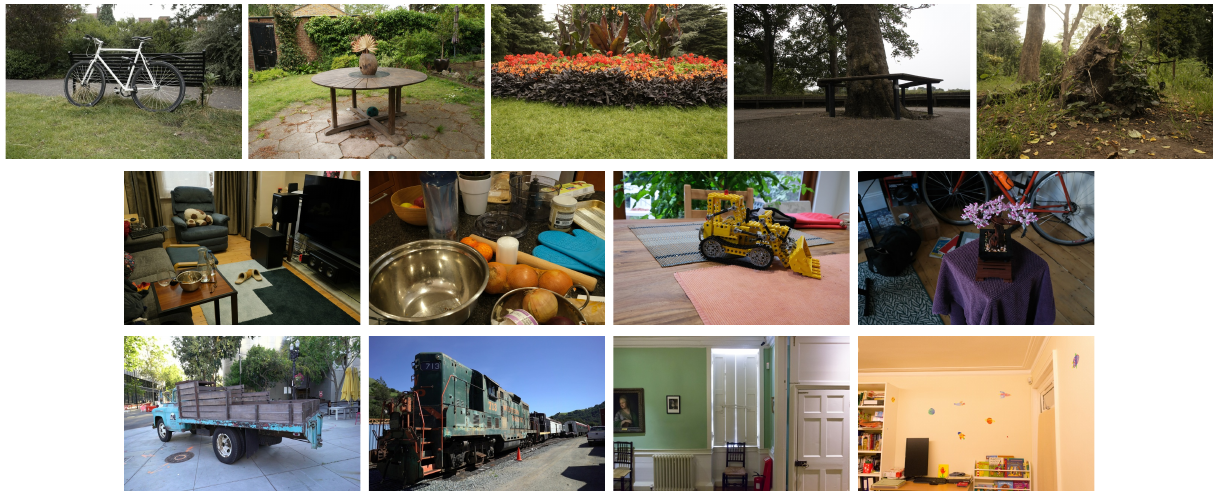} 
\caption{Scenes from three real-world datasets are used for our experimental evaluation.}
\label{fig:datasets} 
\end{figure} 

As shown in Fig.~\ref{fig:datasets}, our experiments are conducted on thirteen scenes from three real-world datasets: 
\begin{itemize} 
    \item Mip-NeRF360~\cite{barron2022mip}: five outdoor scenes, including ``bicycle", ``flowers", ``garden", ``stump", and ``treehill", four indoor scenes, including ``room", ``counter", ``kitchen", and ``bonsai".

    \item  Tanks\&Temples~\cite{knapitsch2017tanks}: two outdoor scenes, including ``truck" and ``train".

    \item Deep Blending~\cite{hedman2018deep}: two indoor scenes, including ``drjohnson" and ``playroom".
    
\end{itemize}

To ensure fair and consistent evaluation, we follow the same preprocessing protocol across these datasets, including scene selection, train/test split, and image resolution, as specified in the official HAC implementation~\cite{chen2024hac}.

\section{More Training and Inference Details}
\subsection{Training Process.}
\begin{figure}[t] 
\centering 
\includegraphics[width=\linewidth]{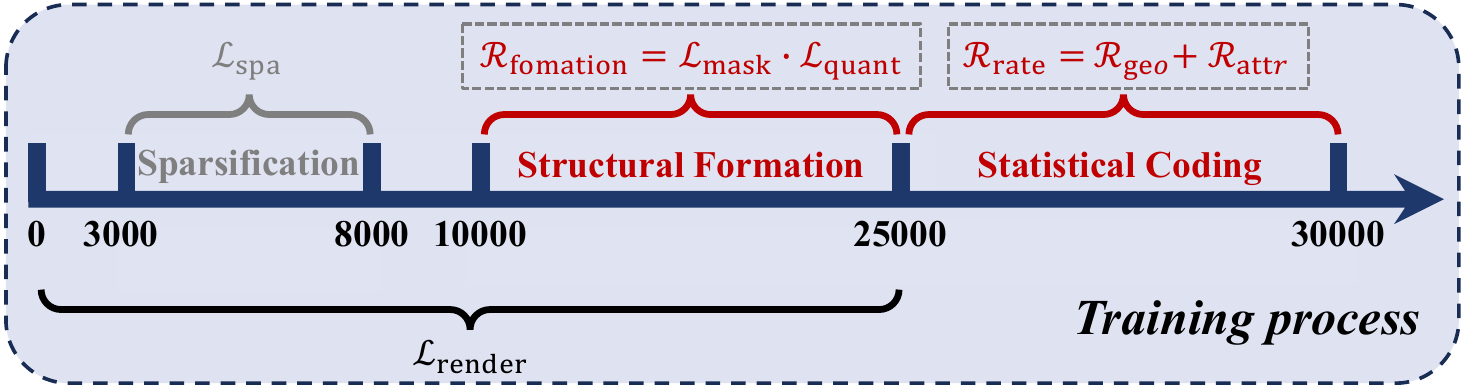} 
\caption{Detailed overall training process of our SpeedyGS. The red fonts indicate those associated with the proposed method.}
\label{fig:training_pipeline} 
\end{figure}

Figure~\ref{fig:training_pipeline} provides an overview of the two-stage training process of SpeedyGS. During the first 10,000 iterations, we follow Mini-Splatting2 to ensure stable optimization of the initial Gaussian representation. Between iterations 3,000 and 8,000, we incorporate GaussianSpa to further improve Gaussian sparsity, as detailed in Sec.~\ref{improving_3dgs_sparsity}.

From iteration 10,000 to 25,000, training enters the structural formation stage. In this stage, we optimize content-aware adaptive quantization and pruning under a proxy-based rate--distortion objective, so as to obtain a compact and coding-friendly 3DGS representation. The learned mask is applied for pruning every 1,000 iterations.

After 25,000 iterations, training proceeds to the statistical coding stage, where the geometry coder and attribute coder are optimized separately on the compacted Gaussian representation produced by the first stage. For efficiency, the geometry coder optimizer is updated every two iterations.

\subsection{Training Objective.}
In the structural formation stage, SpeedyGS minimizes a rate--distortion objective:
\begin{equation}
\mathcal{L}_{\text{total}}
=
\mathcal{D}_{\text{render}}
+
\lambda \cdot \mathcal{R}_{\text{formation}},
\end{equation}
where $\mathcal{D}_{\text{render}}$ denotes the rendering distortion following vanilla 3DGS~\cite{kerbl20233d}, implemented as the combination of L1 loss and SSIM loss. $\mathcal{R}_{\text{formation}}$ denotes the structural formation cost, which is realized by the proposed lightweight rate proxy that jointly accounts for Gaussian precision and Gaussian count through adaptive quantization and pruning.

After obtaining the compacted Gaussian representation, we optimize the statistical coding stage by minimizing the coding rate
\begin{equation}
\mathcal{R}_{\mathrm{coding}}
=
\mathcal{R}_{\mathrm{geo}}
+
\mathcal{R}_{\mathrm{attr}},
\end{equation}
where $\mathcal{R}_{\mathrm{geo}}$ and $\mathcal{R}_{\mathrm{attr}}$ denote the rates estimated by the geometry coder and attribute coder, respectively.

\begin{figure}[t]
\centering 
\includegraphics[width=0.7\linewidth]{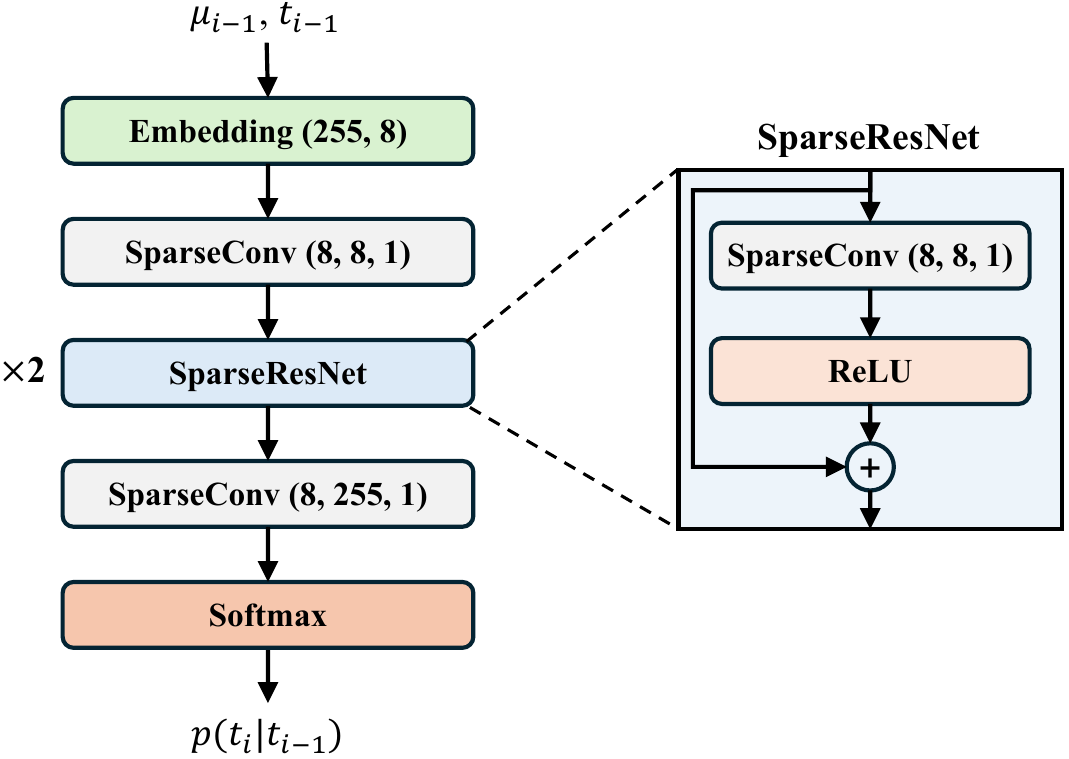}
\caption{The structure of geometry occupancy coding.}
\label{fig:geometry_coding_framework}
\end{figure}

\begin{figure}[t]
\centering 
\includegraphics[width=\linewidth]{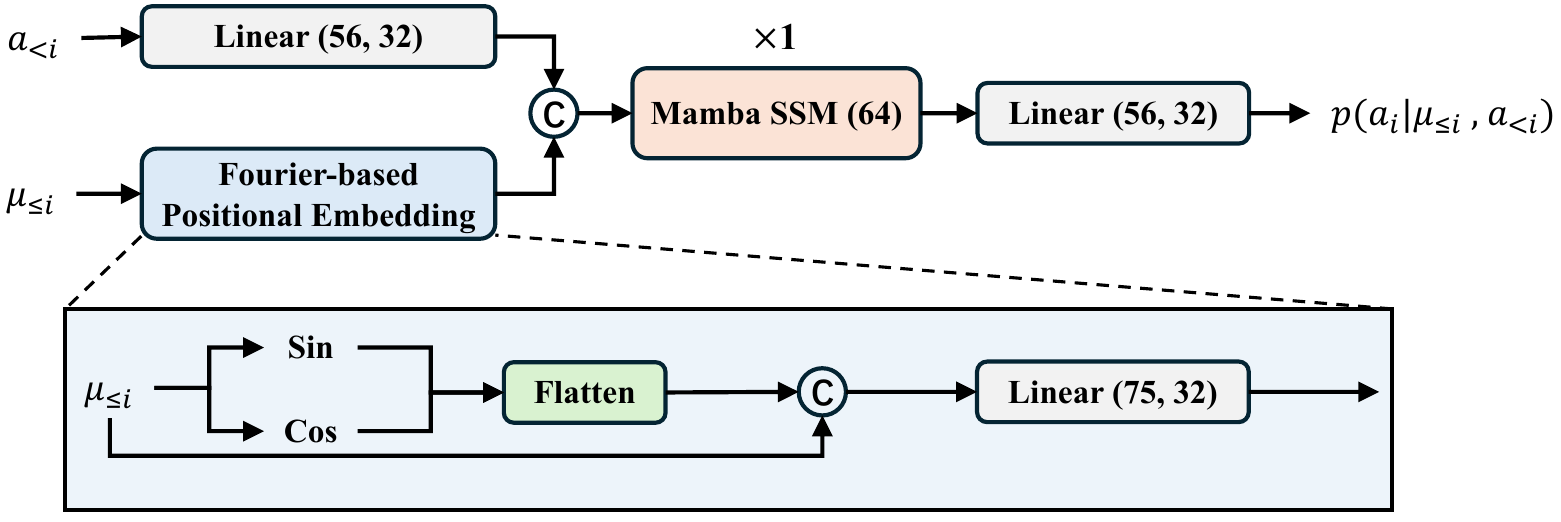}
\caption{The structure of attribute occupancy coding.}
\label{fig:attribute_coding_framework}
\end{figure}

\subsection{Improving 3DGS Sparsity}
\label{improving_3dgs_sparsity}
Since structural formation in SpeedyGS jointly regulates Gaussian density and precision, improving the sparsity of the initial 3DGS representation is beneficial to the subsequent optimization. Our baseline, Mini-Splatting2~\cite{fang2025efficient}, reduces redundancy in vanilla 3DGS via visibility-driven culling, systematically removing Gaussians that contribute negligibly across viewpoints. However, because this removal relies on fixed thresholds, a small number of low-importance Gaussians still remain.

\begin{figure}[t] 
\centering 
\includegraphics[width=\linewidth]{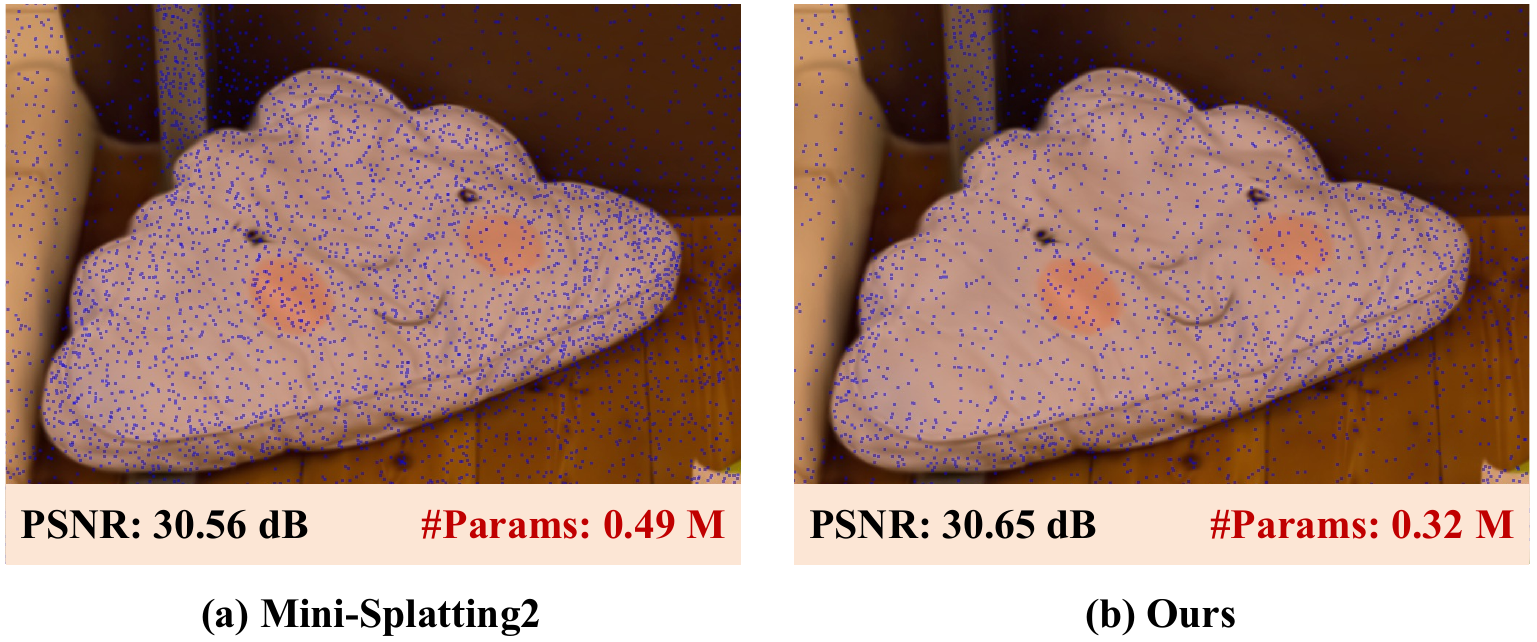} 
\caption{Improving 3DGS Sparsity. Gaussian centers visualized as blue points.}
\label{fig:gaussianSpa} 
\end{figure}

To obtain a more compact initial representation, we further incorporate the sparsification principle of GaussianSpa~\cite{zhang2025gaussianspa}. Specifically, we introduce a dedicated opacity optimization step immediately before visibility culling, which helps drive low-importance Gaussians toward removal. As illustrated in Fig.~\ref{fig:gaussianSpa}, this step substantially reduces redundant Gaussians, lowering the count from 0.49 million to 0.32 million, while yielding a more compact and visually accurate representation. Quantitative results are reported in Table~\ref{tab:gaussianSpa}.

\begin{table*}[htbp]
\centering
\caption{Effect of incorporating GaussianSpa~\cite{zhang2025gaussianspa} to improve sparsity.}
\label{tab:gaussianSpa}
\setlength{\tabcolsep}{4pt}
\renewcommand{\arraystretch}{1.15}
\resizebox{\linewidth}{!}{
\begin{tabular}{lcccccccccccc}
\toprule
\multirow{2}{*}{\textbf{Method}} 
& \multicolumn{4}{c}{\textbf{Mip-NeRF360}} 
& \multicolumn{4}{c}{\textbf{Tanks\&Temples}} 
& \multicolumn{4}{c}{\textbf{DeepBlending}} \\
\cmidrule(lr){2-5}\cmidrule(lr){6-9}\cmidrule(lr){10-13}
& SSIM $\uparrow$ & PSNR $\uparrow$ & LPIPS $\downarrow$ & \#Params (M) $\downarrow$
& SSIM $\uparrow$ & PSNR $\uparrow$ & LPIPS $\downarrow$ & \#Params (M) $\downarrow$
& SSIM $\uparrow$ & PSNR $\uparrow$ & LPIPS $\downarrow$ & \#Params (M) $\downarrow$ \\
\midrule
3DGS              & 0.8130 & 27.49 & 0.222 & 3.16 & 0.8440 & 23.69 & 0.178 & 1.83 & 0.8990 & 29.42 & 0.247 & 2.82 \\
Mini-Splatting2   & 0.8210 & 27.43 & 0.214 & 0.67 & 0.8405 & 23.13 & 0.186 & 0.35 & 0.9120 & 30.04 & 0.240 & 0.65 \\
Ours     & 0.8206 & 27.44 & 0.215 & 0.65 & 0.8409 & 23.16 & 0.187 & 0.34 & 0.9123 & 30.17 & 0.246 & 0.44 \\
\bottomrule
\end{tabular}}
\end{table*}

\section{Bridging Gaussian and Point Cloud}
\begin{figure}[t]
\centering 
\includegraphics[width=\linewidth]{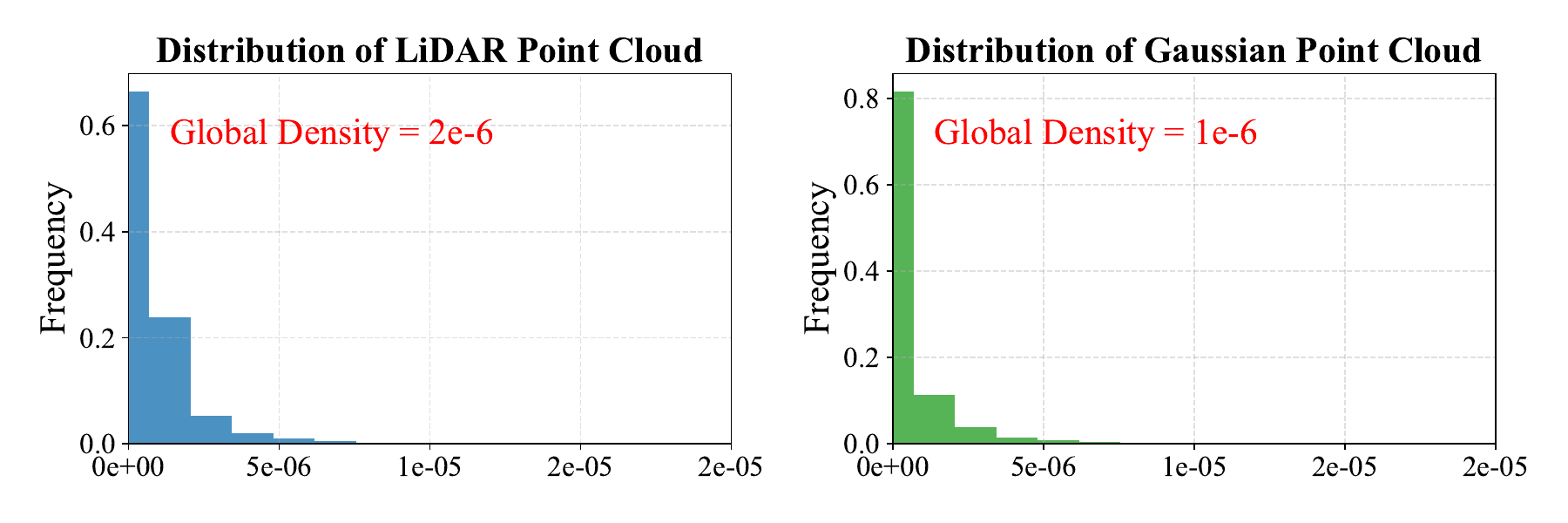}
\caption{Density comparison between LiDAR point cloud and Gaussian point cloud using density analysis. At the same 18-bit geometry precision, the Gaussian point cloud is close to the LiDAR point cloud.}
\label{fig:density}
\end{figure}

Gaussians can essentially be regarded as point clouds enriched with attributes such as color, opacity, scale, and rotation. From a geometric perspective, they share notable similarities with LiDAR point clouds. As shown in Fig.~\ref{fig:density}, a density analysis indicates that the spatial distribution of Gaussian point clouds closely resembles that of LiDAR point clouds when using the same 18-bit coordinate precision. Specifically, the global density of the Gaussian point cloud is calculated to be 2e-6, which is comparable to the LiDAR point cloud's density of 1e-6. Both types of point clouds exhibit sparse and non-uniform structures, with most regions showing very low local density. Therefore, existing LiDAR point cloud compression techniques, which are designed to leverage the sparsity and structural characteristics of such data, can provide valuable guidance for developing effective geometry compression strategies in 3DGS.

\section{Coding Algorithm Analysis}
\subsection{Structure Details of SpeedyGS}

This subsection details the two components of the statistical coding stage, namely geometry coding and attribute coding, as illustrated in Fig.~\ref{fig:geometry_coding_framework} and Fig.~\ref{fig:attribute_coding_framework}.

For geometry coding, we represent Gaussian coordinates using sparse octree tokens and model them with a sparse convolutional neural network~\cite{choy20194d}. The previous octree token is embedded into a low-dimensional feature, processed by two SparseResNet blocks, and then projected by a sparse convolution and a Softmax layer to predict the probability of each 8-bit occupancy token.

For attribute coding, we employ a compact causal Mamba SSM in a local autoregressive manner. Each attribute token is linearly embedded and concatenated with a Fourier positional embedding derived from its geometric coordinates. The resulting feature is then passed through a single Mamba block and a linear prediction head to model the conditional distribution of the next attribute token.

\subsection{Compressed Data Precision Analysis}

\begin{figure}[t]
\centering 
\includegraphics[width=\linewidth]{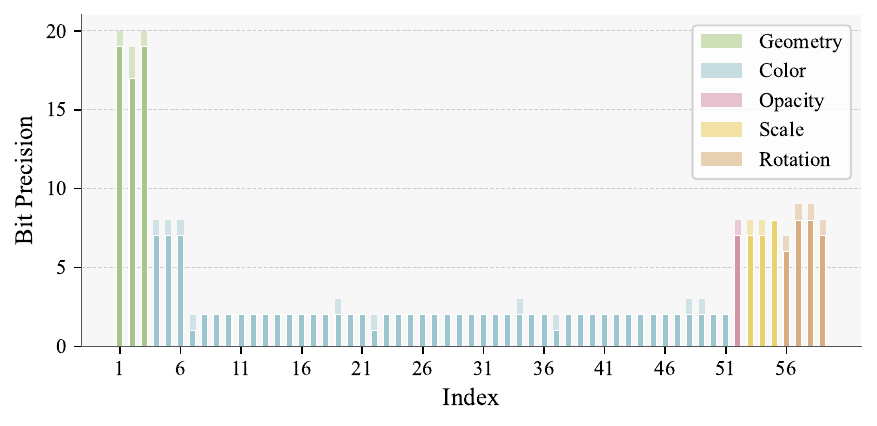}
\caption{Bit precision of each Gaussian channel after structural formation. 
Solid bars show the low-rate ($\lambda=2e-4$), and translucent bars indicate the high-rate ($\lambda=2.5e-5$). Scene: bicycle.}
\label{fig:precision_analysis}
\end{figure}

After the structural formation stage, we obtain the compacted Gaussian with fewer primitives and lower numerical precision. To better understand the structural formation process, we further analyze the resulting bit precision of the optimized Gaussians, revealing how the model organizes and quantizes different channels during structural formation. As shown in Fig.~\ref{fig:precision_analysis}, the bit precision distribution exhibits a clear structure: geometry channels retain high precision, while color precision is concentrated in the first three channels, with the remaining 45 color channels compressed to very low bit widths. Opacity, scale, and rotation occupy a narrow mid-precision band.

\subsection{Bitstream Composition}
\begin{figure}[t] 
\centering 
\includegraphics[width=0.99\linewidth]{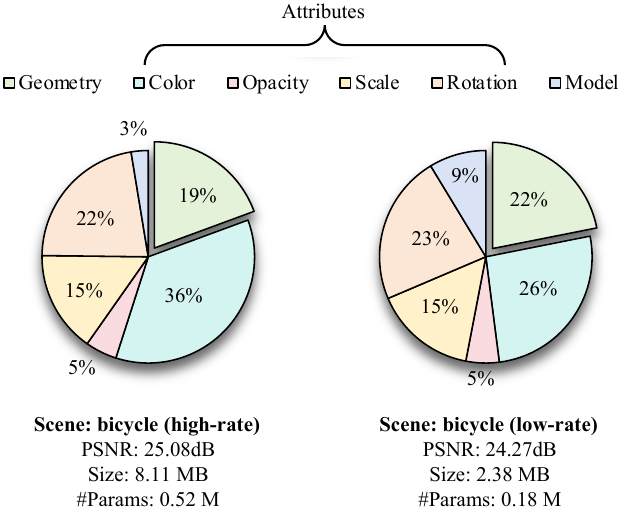} 
\caption{Storage size of each component in low-rate ($\lambda=2e-4$) and high-rate ($\lambda=2.5e-5$). Scene: bicycle.}
\label{fig:bit_allocation} 
\end{figure}

To better understand how the statistical coding stage further reduces redundancy on top of the compacted Gaussian representation produced by structural formation, we analyze the bitstream generated by SpeedyGS, which consists of three main parts.
\begin{itemize}
    \item Geometry: coordinates $\boldsymbol\mu\in\mathbb R^3$.
    \item Attributes: color $\boldsymbol c\in\mathbb R^{48}$ (including base color $\boldsymbol c_{\text{base}}\in\mathbb R^3$ and Spherical Harmonic coefficients $\boldsymbol c_{\text{SH}}\in\mathbb R^{45}$), opacity $\boldsymbol \alpha\in\mathbb R$, scale $\boldsymbol s\in\mathbb R^3$, and rotation $\boldsymbol r\in\mathbb R^4$.
    \item Model parameters: parameters of the geometry coder and attribute coder, using LZMA~\cite{pavlov2009lzma} lossless coding.
\end{itemize}
Figure~\ref{fig:bit_allocation} shows the size of each composition. Using the ``bicycle" (high-rate) as an example, the specific bit allocations are 1.57 MB for geometry, 6.33 MB (2.88 MB for color, 0.39 MB for opacity, 1.25 MB for scale, 1.81 MB for rotation) for attributes, and 0.21 MB for model parameters. As the scene is simplified, both the size and relative proportions of each component change accordingly.

\begin{figure}[t] 
\centering 
\includegraphics[width=0.99\linewidth]{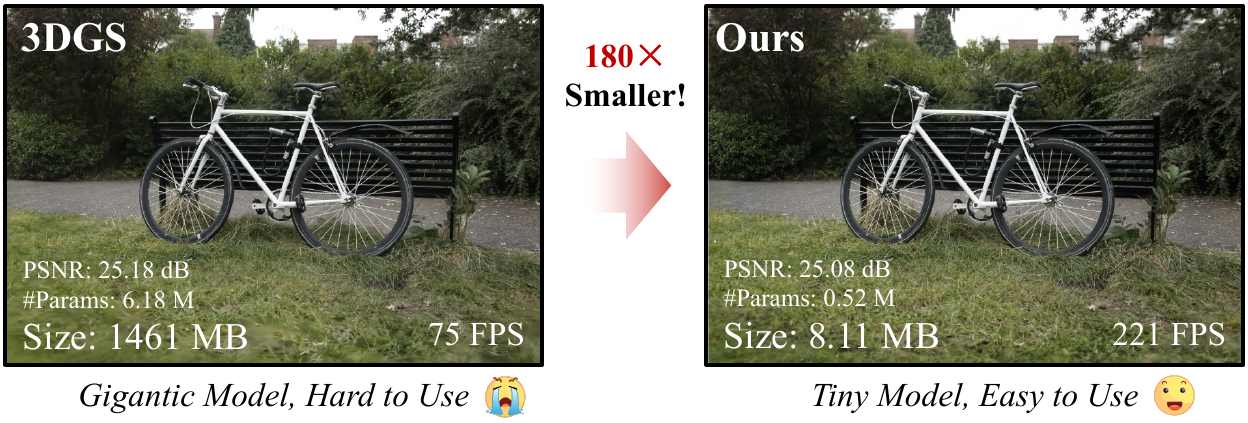} 
\caption{Compared to vanilla 3DGS, SpeedyGS achieves a 180× compression ratio while maintaining the same image quality for the testing scene. Scene: bicycle.}
\label{fig:vs_3dgs} 
\end{figure}

\subsection{Complexity Analysis}
Table~\ref{tab:decoding_complexity} breaks down the decoding time of SpeedyGS into geometry and attribute decoding, each further divided into neural network (NN) inference and arithmetic coding (AC). In both settings, attribute decoding is the main bottleneck. For high-rate ($\lambda=2.5e-5$), the full model takes 3.27\,s in total, where geometry decoding requires only 0.24\,s while attribute decoding takes 3.03\,s (92.7\%). For low-rate ($\lambda=2e-4$), the total decoding time is 2.06\,s, including 0.17\,s for geometry and 1.89\,s for attributes (91.7\%). Moreover, the dominant cost mainly comes from AC rather than NN inference, indicating that the network modules are lightweight.

The Ours-Fast variant significantly reduces the attribute decoding cost, decreasing the total decoding time from 3.27\,s to 2.14\,s at high-rate and from 2.06\,s to 0.78\,s at low-rate. In addition, Ours-UltraFast nearly removes decoding latency altogether.

\begin{table}[htbp]
\centering
\caption{Breakdown of decoding time into neural network (NN) inference and arithmetic coding (AC) for geometry and attribute decoding under different configurations.}
\label{tab:decoding_complexity}
\resizebox{\linewidth}{!}{
\begin{tabular}{c c c c c c}
\toprule
$\lambda$ & Category & Component & Ours & Ours-Fast & Ours-UltraFast \\
\midrule

\multirow{7}{*}{2.5e-5}
& \multirow{3}{*}{Geometry}
& NN (s)    & 0.10 & 0.10 & -- \\
&  & AC (s)    & 0.14 & 0.14 & -- \\
&  & Total (s) & 0.24 & 0.24 & 0.00005 \\
\cmidrule(lr){2-6}
& \multirow{3}{*}{Attribute}
& NN (s)    & 0.24 & 0.06 & -- \\
&  & AC (s)    & 2.79 & 1.84 & -- \\
&  & Total (s) & 3.03 & 1.90 & 0.00020 \\
\cmidrule(lr){2-6}
& \multicolumn{2}{c}{Overall Total (s)} & 3.27 & 2.14 & 0.00025 \\
\midrule

\multirow{7}{*}{2e-4}
& \multirow{3}{*}{Geometry}
& NN (s)    & 0.05 & 0.05 & -- \\
&  & AC (s)    & 0.12 & 0.12 & -- \\
&  & Total (s) & 0.17 & 0.17 & 0.00005 \\
\cmidrule(lr){2-6}
& \multirow{3}{*}{Attribute}
& NN (s)    & 0.20 & 0.05 & -- \\
&  & AC (s)    & 1.69 & 0.56 & -- \\
&  & Total (s) & 1.89 & 0.61 & 0.00020 \\
\cmidrule(lr){2-6}
& \multicolumn{2}{c}{Overall Total (s)} & 2.06 & 0.78 & 0.00025 \\
\bottomrule
\end{tabular}}
\end{table}

\section{More Experimental Results}
\subsection{Additional Quantitative Results}
\begin{figure*}[htbp]
  \centering
\subfloat[]
{\includegraphics[width=0.98\linewidth]{Figures/Exp/rd_mipnerf360.pdf}}\\
\subfloat[]
{\includegraphics[width=0.98\linewidth]{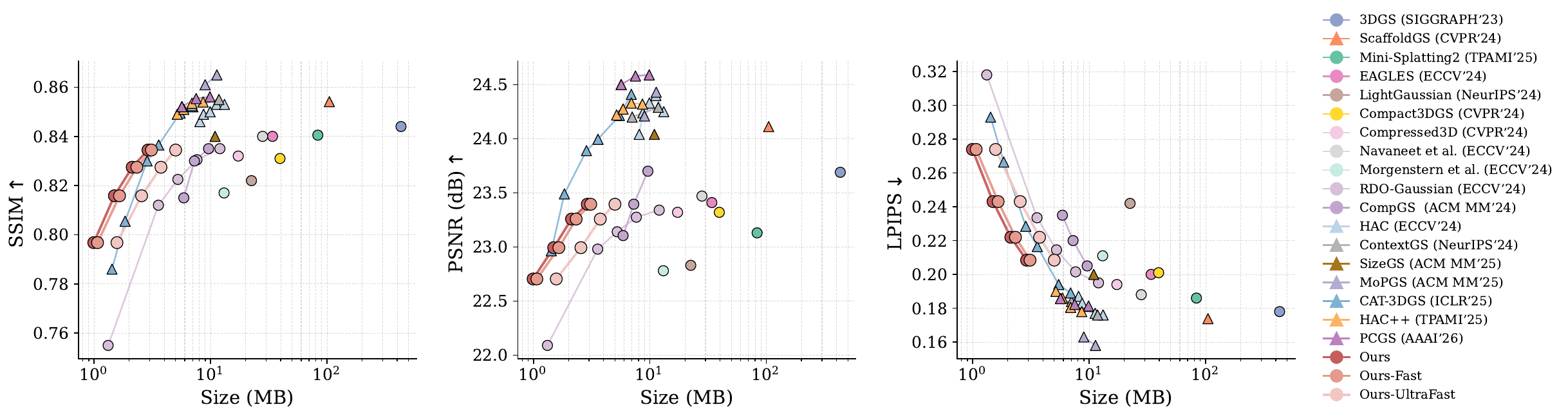}}\\
\subfloat[]
{\includegraphics[width=0.98\linewidth]{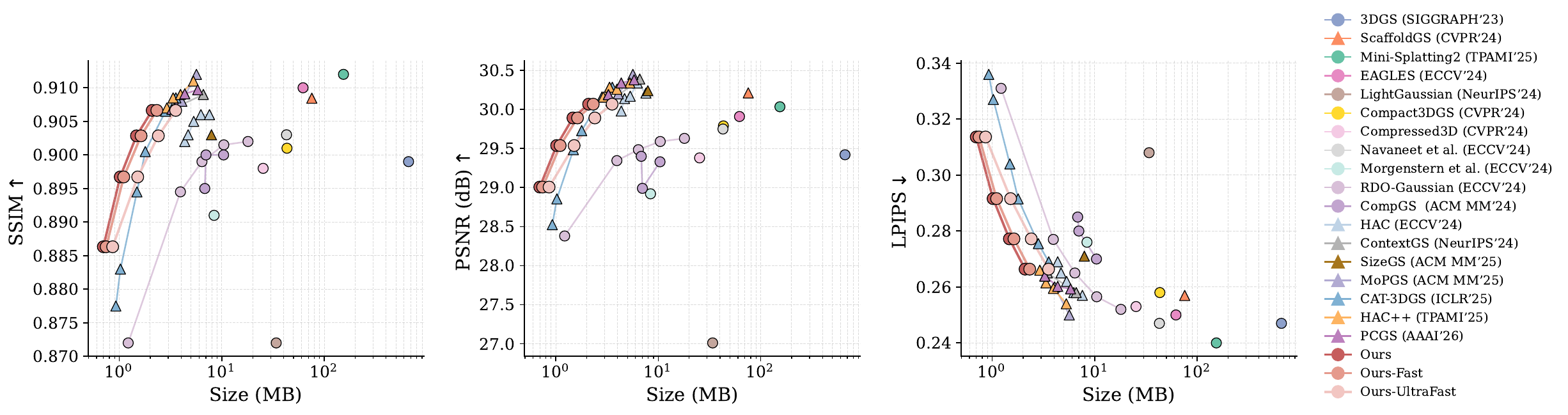}}\\
  \caption{Rate-distortion curves for quantitative comparisons. (a) Mip-NeRF360. (b) Tanks\&Temples (c) DeepBlending. We obtain multiple R-D points by setting different $\lambda$. The proposed method is based on the improved work Mini-Splatting2~\cite{fang2025efficient} of vanilla 3DGS, so the upper limit of fidelity will be constrained by the baseline.}
  \label{fig:rd_curve}
\end{figure*}

We compare SpeedyGS with previous 3DGS compression methods, including vanilla 3DGS~\cite{kerbl20233d}, Scaffold-GS~\cite{lu2024scaffold}, nine 3DGS-based approaches~\cite{fang2025efficient, girish2024eagles,fan2024lightgaussian,lee2024compact,niedermayr2024compressed,navaneet2024compgs,morgenstern2024compact,wang2024end,liu2024compgs}, and seven Scaffold-GS-based approaches~\cite{chen2024hac,chen2025hac++,chen2026pcgs,wang2024contextgs,zhan2025cat3dgs,liu20253d,xie2025sizegs}. Table~\ref{tab:more_quantitative_results} reports SSIM, PSNR, LPIPS, and size on Mip-NeRF360~\cite{barron2022mip}, Tanks\&Temples~\cite{knapitsch2017tanks}, and DeepBlending~\cite{hedman2018deep}. The corresponding rate-distortion curves are shown in Fig.~\ref{fig:rd_curve}.

On Mip-NeRF360, SpeedyGS achieves about a 110$\times$ reduction in size relative to vanilla 3DGS at similar fidelity. A representative example is the reduction of a 1461 MB scene to 8.11 MB, as shown in Fig.~\ref{fig:vs_3dgs}. SpeedyGS also achieves more than a 23$\times$ improvement in compression ratio over Mini-Splatting2. Compared with CAT-3DGS, SpeedyGS and SpeedyGS-Fast achieve 50.52\% and 44.81\% BD-Rate gains on SSIM, respectively. Compared with other 3DGS-based methods such as RDO-Gaussian and CompGS, SpeedyGS yields 3.60$\times$ and 2.53$\times$ higher compression ratios, respectively.

On Tanks\&Temples and DeepBlending, SpeedyGS also demonstrates clear compression advantages. Specifically, it reduces size by factors of 149$\times$ and 318$\times$ relative to vanilla 3DGS on these two datasets, respectively. In addition, SpeedyGS achieves a 29.86\% BD-Rate gain over CAT-3DGS on PSNR on DeepBlending.

Overall, these results further support the main claim of the paper: by decoupling structural formation and statistical coding, SpeedyGS achieves a favorable balance among compression performance, optimization efficiency, decoding latency, and rendering speed. Detailed per-scene results are reported in Table~\ref{tab:each_scene_results_mipnerf360}, Table~\ref{tab:each_scene_results_tandt}, and Table~\ref{tab:each_scene_results_deepblending}.

\subsection{Additional Qualitative Results}
\begin{figure*}[t]
\centering 
\includegraphics[width=0.98\linewidth]{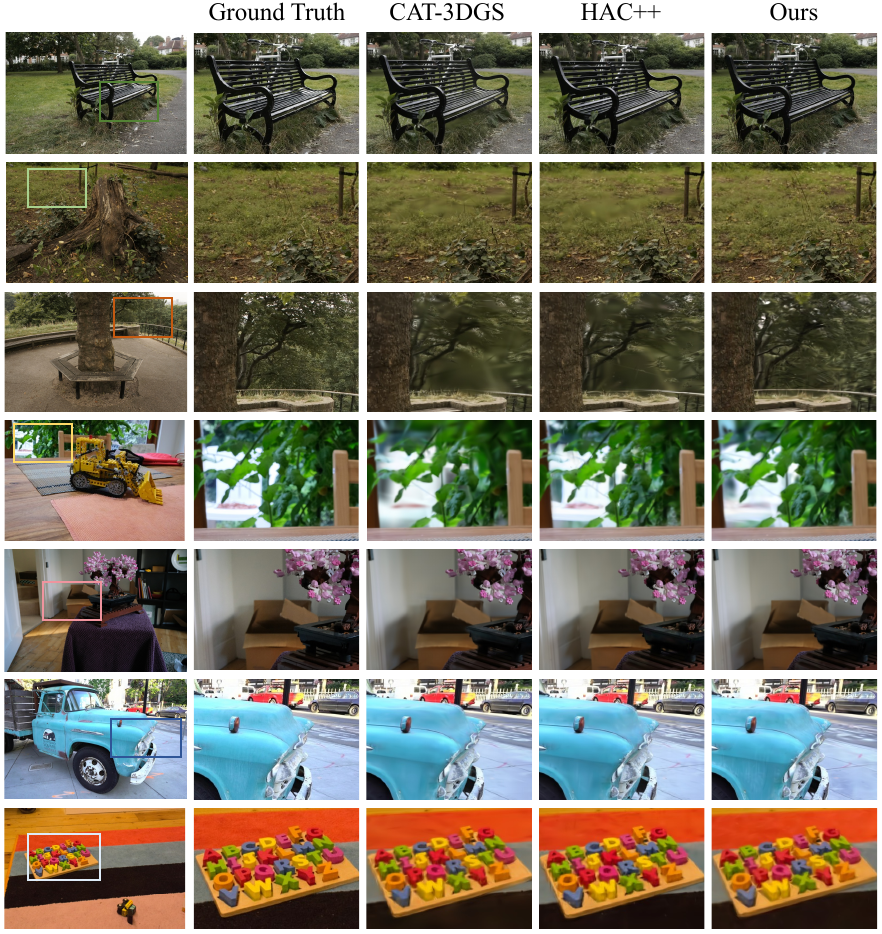} 
\caption{Qualitative Results. Best viewed zoomed in.}
\label{fig:more_quantitative_results} 
\end{figure*}

Figure~\ref{fig:more_quantitative_results} presents additional rendering comparisons, evaluating the visual quality of SpeedyGS versus CAT‑3DGS~\cite{zhan2025cat3dgs}, and HAC++~\cite{chen2025hac++} across a variety of scenes.

\begin{table*}[htbp]
\centering
\caption{Quantitative comparison of \textbf{SpeedyGS} against prior methods on Mip-NeRF360, Tanks\&Temples, and DeepBlending.}
\label{tab:more_quantitative_results}
\setlength{\tabcolsep}{3.2pt}
\renewcommand{\arraystretch}{1.15}
\resizebox{\linewidth}{!}{
\begin{tabular}{lcccccccccccc}
\toprule
\multirow{2}{*}{\textbf{Method}} 
& \multicolumn{4}{c}{\textbf{Mip-NeRF360}} 
& \multicolumn{4}{c}{\textbf{Tanks\&Temples}} 
& \multicolumn{4}{c}{\textbf{DeepBlending}} \\
\cmidrule(lr){2-5}\cmidrule(lr){6-9}\cmidrule(lr){10-13}
& SSIM $\uparrow$ & PSNR $\uparrow$ & LPIPS $\downarrow$ & Size (MB) $\downarrow$
& SSIM $\uparrow$ & PSNR $\uparrow$ & LPIPS $\downarrow$ & Size (MB) $\downarrow$
& SSIM $\uparrow$ & PSNR $\uparrow$ & LPIPS $\downarrow$ & Size (MB) $\downarrow$ \\
\midrule

3DGS (SIGGRAPH'23)                     & 0.813 & 27.49 & 0.222 & 744.70 & 0.844 & 23.69 & 0.178 & 431.00 & 0.899 & 29.42 & 0.247 & 663.90 \\
Scaffold-GS (CVPR'24)                 & 0.812 & 27.81 & 0.228 & 227.95 & 0.854 & 24.11 & 0.174 & 104.60  & 0.908 & 30.21 & 0.257 & 75.52  \\
\midrule

Mini-Splatting2 (TPAMI'25)                       & 0.821 & 27.43 & 0.214 & 159.18 & 0.841 & 23.13 & 0.186 & 83.26  & 0.912 & 30.04 & 0.240 & 153.80 \\
EAGLES (ECCV'24)                      & 0.808 & 27.15 & 0.238 & 68.89  & 0.840 & 23.41 & 0.200 & 34.00  & 0.910 & 29.91 & 0.250 & 62.00  \\
Navaneet et al. (ECCV'24)             & 0.808 & 27.16 & 0.228 & 50.30  & 0.840 & 23.47 & 0.188 & 27.97  & 0.903 & 29.75 & 0.247 & 42.77  \\
Morgenstern et al. (ECCV'24)          & 0.772 & 26.01 & 0.259 & 23.90  & 0.817 & 22.78 & 0.211 & 13.05  & 0.891 & 28.92 & 0.276 & 8.40   \\
LightGaussian (NeurIPS'24)            & 0.799 & 27.00 & 0.249 & 44.54  & 0.822 & 22.83 & 0.242 & 22.43  & 0.872 & 27.01 & 0.308 & 33.94  \\
Compact3DGS (CVPR'24)                 & 0.798 & 27.08 & 0.247 & 48.80  & 0.831 & 23.32 & 0.201 & 39.43  & 0.901 & 29.79 & 0.258 & 43.21  \\
Compressed3D (CVPR'24)                & 0.801 & 26.98 & 0.238 & 28.80  & 0.832 & 23.32 & 0.194 & 17.28  & 0.898 & 29.38 & 0.253 & 25.30  \\
\midrule

RDO-Gaussian (ECCV'24, high-rate)     & 0.802 & 27.05 & 0.239 & 23.46  & 0.835 & 23.34 & 0.195 & 12.02  & 0.902 & 29.63 & 0.252 & 18.00  \\
RDO-Gaussian (ECCV'24, low-rate)      & 0.683 & 24.43 & 0.406 & 1.71   & 0.755 & 22.09 & 0.318 & 1.32   & 0.872 & 28.38 & 0.331 & 1.22   \\
CompGS (ACM MM'24, high-rate)         & 0.802 & 27.26 & 0.239 & 16.50  & 0.835 & 23.70 & 0.205 & 9.61   & 0.900 & 29.33 & 0.270 & 10.40  \\
CompGS (ACM MM'24, low-rate)          & 0.791 & 26.79 & 0.258 & 11.00  & 0.815 & 23.11 & 0.235 & 5.89   & 0.900 & 28.99 & 0.280 & 7.00   \\
\midrule

HAC (ECCV'24, high-rate)              & 0.811 & 27.77 & 0.230 & 21.87  & 0.853 & 24.40 & 0.177 & 11.24  & 0.906 & 30.34 & 0.258 & 6.35   \\
HAC (ECCV'24, low-rate)               & 0.807 & 27.53 & 0.238 & 15.26  & 0.846 & 24.04 & 0.187 & 8.10   & 0.902 & 29.98 & 0.269 & 4.35   \\
ContextGS (NeurIPS'24, high-rate)     & 0.811 & 27.75 & 0.231 & 18.41  & 0.855 & 24.29 & 0.176 & 11.80  & 0.909 & 30.39 & 0.258 & 6.60   \\
ContextGS (NeurIPS'24, low-rate)      & 0.808 & 27.62 & 0.237 & 12.68  & 0.852 & 24.20 & 0.184 & 7.05   & 0.907 & 30.11 & 0.265 & 3.45   \\
SizeGS (ACM MM'25)          & 0.806  & 27.48  & 0.240 & 18.17  & 0.840 & 24.04 & 0.200 & 10.93  & 0.903 & 30.24 & 0.271 & 7.92   \\
MoPGS (ACM MM'25, high-rate)          & 0.811 & 27.89 & 0.227 & 21.89  & 0.865 & 24.43 & 0.158 & 11.33  & 0.912 & 30.45 & 0.250 & 5.65   \\
MoPGS (ACM MM'25, low-rate)           & 0.808 & 27.68 & 0.234 & 15.64  & 0.861 & 24.21 & 0.163 & 8.98   & 0.908 & 30.20 & 0.260 & 4.07   \\
CAT-3DGS (ICLR'25, high-rate)         & 0.809 & 27.77 & 0.241 & 12.35  & 0.853 & 24.41 & 0.189 & 6.93   & 0.909 & 30.29 & 0.269 & 3.56   \\
CAT-3DGS (ICLR'25, low-rate)          & 0.730 & 25.82 & 0.362 & 1.72   & 0.786 & 22.97 & 0.293 & 1.42   & 0.878 & 28.53 & 0.336 & 0.93   \\
HAC++ (TPAMI'25, high-rate)           & 0.811 & 27.82 & 0.231 & 18.48  & 0.854 & 24.32 & 0.178 & 8.63   & 0.911 & 30.34 & 0.254 & 5.28   \\
HAC++ (TPAMI'25, low-rate)            & 0.803 & 27.60 & 0.253 & 8.34   & 0.849 & 24.22 & 0.190 & 5.18   & 0.907 & 30.16 & 0.266 & 2.91   \\
PCGS (AAAI'26, high-rate)             & 0.811 & 27.96 & 0.236 & 24.74  & 0.856 & 24.59 & 0.181 & 9.87   & 0.910 & 30.38 & 0.259 & 5.81   \\
PCGS (AAAI'26, low-rate)              & 0.808 & 27.82 & 0.240 & 12.05  & 0.852 & 24.50 & 0.186 & 5.67   & 0.907 & 30.19 & 0.264 & 3.26   \\
\midrule

Ours (high-rate)            & 0.815 & 27.32 & 0.228 & 6.51  & 0.834 & 23.39 & 0.208 & 2.90 & 0.907 & 30.07 & 0.266 & 2.09 \\
Ours (low-rate)             & 0.777 & 26.47 & 0.300 & 2.01  & 0.797 & 22.70 & 0.274 & 0.99 & 0.886 & 29.01 & 0.314 & 0.70 \\
Ours-Fast (high-rate)      & 0.815 & 27.32 & 0.228 & 7.23  & 0.834 & 23.39 & 0.208 & 3.11 & 0.907 & 30.07 & 0.266 & 2.32 \\
Ours-Fast (low-rate)       & 0.777 & 26.47 & 0.300 & 2.22  & 0.797 & 22.70 & 0.274 & 1.07 & 0.886 & 29.01 & 0.314 & 0.74 \\
Ours-UltraFast (high-rate) & 0.815 & 27.32 & 0.228 & 11.95 & 0.834 & 23.39 & 0.208 & 5.01 & 0.907 & 30.07 & 0.266 & 3.55 \\
Ours-UltraFast (low-rate)  & 0.777 & 26.47 & 0.300 & 3.75  & 0.797 & 22.70 & 0.274 & 1.57 & 0.886 & 29.01 & 0.314 & 0.86 \\
\bottomrule
\end{tabular}
}
\end{table*}

\begin{table*}[t]
\centering
\scriptsize
\setlength{\tabcolsep}{4pt}
\renewcommand{\arraystretch}{1.15}
\caption{Per-scene results of SpeedyGS and its Fast and UltraFast variants on the Mip-NeRF360 dataset. Note that the decoding time of the UltraFast version is nearly zero.}
\label{tab:each_scene_results_mipnerf360}
\resizebox{\textwidth}{!}{
\begin{tabular}{c|c|ccc|ccc|ccc|cc|c}
\toprule
\multirow{2}{*}{$\lambda$} & \multirow{2}{*}{Dataset} & \multicolumn{3}{c|}{Fidelity} & \multicolumn{3}{c|}{Size $\downarrow$} & \multicolumn{3}{c|}{Training Time $\downarrow$} & \multicolumn{2}{c|}{Decoding Time $\downarrow$} & \multirow{2}{*}{\makecell{\#Params\\(M)}} \\
& & SSIM $\uparrow$ & PSNR $\uparrow$ & LPIPS $\downarrow$ & Ours & Fast & UltraFast & Ours & Fast & UltraFast & Ours & Fast &\\
\midrule

\multirow{10}{*}{2e-4}
& bicycle  & 0.6903 & 24.27 & 0.3483 & 2.38 & 2.67 & 4.62 & 496 & 448 & 257 & 2.09 & 0.83 & 0.18 \\
& bonsai   & 0.9192 & 29.96 & 0.2314 & 1.26 & 1.39 & 2.12 & 449 & 448 & 268 & 1.93 & 0.66 & 0.08 \\
& counter  & 0.8770 & 27.56 & 0.2514 & 1.28 & 1.40 & 1.93 & 473 & 438 & 269 & 1.95 & 0.66 & 0.07 \\
& flowers  & 0.5514 & 21.07 & 0.4219 & 2.48 & 2.74 & 4.80 & 496 & 455 & 256 & 2.28 & 0.94 & 0.19 \\
& garden   & 0.7906 & 25.92 & 0.2505 & 3.41 & 3.79 & 7.16 & 560 & 516 & 303 & 2.22 & 0.98 & 0.28 \\
& kitchen  & 0.9018 & 29.87 & 0.1772 & 1.68 & 1.82 & 2.83 & 495 & 479 & 290 & 2.04 & 0.72 & 0.11 \\
& room     & 0.8981 & 30.48 & 0.2679 & 0.83 & 0.90 & 1.25 & 432 & 409 & 247 & 1.86 & 0.53 & 0.05 \\
& stump    & 0.7506 & 26.35 & 0.3176 & 2.72 & 3.03 & 5.14 & 480 & 450 & 254 & 2.10 & 0.99 & 0.19 \\
& treehill & 0.6140 & 22.74 & 0.4352 & 2.06 & 2.25 & 3.90 & 446 & 431 & 241 & 2.09 & 0.75 & 0.15 \\
& \textbf{AVG}      & 0.7770 & 26.47 & 0.3002 & 2.01 & 2.22 & 3.75 & 481 & 453 & 265 & 2.06 & 0.78 & 0.14 \\
\midrule

\multirow{10}{*}{1e-4}
& bicycle  & 0.7255 & 24.68 & 0.3022 & 3.94 & 4.38 & 7.75 & 499 & 490 & 282 & 2.41 & 1.13 & 0.29 \\
& bonsai   & 0.9318 & 30.75 & 0.2062 & 1.89 & 2.10 & 3.45 & 495 & 480 & 282 & 2.01 & 0.83 & 0.13 \\
& counter  & 0.8917 & 28.04 & 0.2247 & 1.94 & 2.19 & 3.18 & 496 & 474 & 285 & 2.03 & 0.73 & 0.11 \\
& flowers  & 0.5763 & 21.28 & 0.3875 & 3.98 & 4.45 & 7.86 & 512 & 488 & 278 & 2.38 & 1.29 & 0.30 \\
& garden   & 0.8187 & 26.40 & 0.1993 & 5.47 & 6.16 & 11.39 & 584 & 569 & 336 & 2.80 & 1.65 & 0.44 \\
& kitchen  & 0.9156 & 30.51 & 0.1490 & 2.57 & 2.82 & 4.57 & 543 & 508 & 304 & 2.18 & 0.91 & 0.17 \\
& room     & 0.9117 & 31.04 & 0.2369 & 1.30 & 1.41 & 2.19 & 468 & 439 & 256 & 2.03 & 0.63 & 0.09 \\
& stump    & 0.7757 & 26.77 & 0.2757 & 4.45 & 5.12 & 8.50 & 519 & 489 & 275 & 2.85 & 1.40 & 0.31 \\
& treehill & 0.6361 & 22.85 & 0.3897 & 3.48 & 3.82 & 6.69 & 494 & 467 & 261 & 2.32 & 1.07 & 0.25 \\
& \textbf{AVG}      & 0.7981 & 26.92 & 0.2635 & 3.23 & 3.60 & 6.18 & 512 & 489 & 284 & 2.33 & 1.07 & 0.23 \\
\midrule

\multirow{10}{*}{5e-5}
& bicycle  & 0.7452 & 24.95 & 0.2699 & 5.95 & 6.70 & 11.33 & 529 & 530 & 302 & 3.00 & 1.70 & 0.42 \\
& bonsai   & 0.9390 & 31.20 & 0.1902 & 2.79 & 3.04 & 5.20 & 498 & 498 & 295 & 2.56 & 1.13 & 0.19 \\
& counter  & 0.9008 & 28.25 & 0.2057 & 2.90 & 3.29 & 4.93 & 501 & 500 & 300 & 2.18 & 0.90 & 0.17 \\
& flowers  & 0.5906 & 21.35 & 0.3661 & 5.93 & 6.44 & 11.22 & 518 & 518 & 297 & 3.10 & 1.74 & 0.41 \\
& garden   & 0.8324 & 26.67 & 0.1729 & 7.91 & 8.96 & 15.59 & 635 & 610 & 361 & 3.40 & 2.43 & 0.59 \\
& kitchen  & 0.9235 & 31.03 & 0.1335 & 3.83 & 4.23 & 6.87 & 533 & 537 & 325 & 2.47 & 1.11 & 0.25 \\
& room     & 0.9201 & 31.35 & 0.2140 & 1.95 & 2.17 & 3.60 & 456 & 469 & 270 & 2.09 & 0.78 & 0.13 \\
& stump    & 0.7894 & 27.00 & 0.2487 & 6.67 & 7.54 & 12.34 & 524 & 527 & 298 & 2.97 & 1.73 & 0.44 \\
& treehill & 0.6472 & 22.91 & 0.3621 & 5.24 & 5.82 & 9.75 & 490 & 498 & 280 & 2.73 & 1.49 & 0.36 \\
& \textbf{AVG}      & 0.8098 & 27.19 & 0.2403 & 4.80 & 5.35 & 8.98 & 520 & 521 & 303 & 2.72 & 1.45 & 0.33 \\
\midrule

\multirow{10}{*}{2.5e-5}
& bicycle  & 0.7550 & 25.08 & 0.2527 & 8.11 & 9.01 & 14.93 & 578 & 548 & 314 & 3.88 & 2.70 & 0.52 \\
& bonsai   & 0.9428 & 31.47 & 0.1808 & 3.87 & 4.30 & 7.34 & 524 & 515 & 306 & 2.94 & 1.64 & 0.25 \\
& counter  & 0.9064 & 28.45 & 0.1946 & 4.08 & 4.54 & 7.08 & 536 & 514 & 308 & 2.54 & 1.24 & 0.23 \\
& flowers  & 0.5961 & 21.37 & 0.3556 & 7.73 & 8.49 & 14.16 & 559 & 532 & 307 & 3.64 & 2.49 & 0.49 \\
& garden   & 0.8391 & 26.82 & 0.1600 & 10.49 & 11.71 & 19.85 & 660 & 633 & 371 & 4.16 & 3.70 & 0.71 \\
& kitchen  & 0.9282 & 31.25 & 0.1254 & 5.35 & 5.84 & 9.60 & 573 & 562 & 338 & 2.90 & 1.55 & 0.34 \\
& room     & 0.9254 & 31.53 & 0.2007 & 2.79 & 3.08 & 5.19 & 493 & 479 & 275 & 2.15 & 0.90 & 0.19 \\
& stump    & 0.7947 & 27.07 & 0.2355 & 9.04 & 10.15 & 16.87 & 563 & 546 & 311 & 3.75 & 2.56 & 0.54 \\
& treehill & 0.6515 & 22.86 & 0.3475 & 7.10 & 7.97 & 12.52 & 542 & 514 & 287 & 3.47 & 2.49 & 0.44 \\
& \textbf{AVG}      & 0.8155 & 27.32 & 0.2281 & 6.51 & 7.23 & 11.95 & 559 & 538 & 313 & 3.27 & 2.14 & 0.41 \\

\bottomrule
\end{tabular}
}
\end{table*}

\begin{table*}[t]
\centering
\scriptsize
\setlength{\tabcolsep}{4pt}
\renewcommand{\arraystretch}{1.15}
\caption{Per-scene results of SpeedyGS on the Tanks\&Temples dataset. Note that the decoding time of the UltraFast version is nearly zero.}
\label{tab:each_scene_results_tandt}
\resizebox{\textwidth}{!}{
\begin{tabular}{c|c|ccc|ccc|ccc|cc|c}
\toprule
\multirow{2}{*}{$\lambda$} & \multirow{2}{*}{Dataset} & \multicolumn{3}{c|}{Fidelity} & \multicolumn{3}{c|}{Size $\downarrow$} & \multicolumn{3}{c|}{Training Time $\downarrow$} & \multicolumn{2}{c|}{Decoding Time $\downarrow$} & \multirow{2}{*}{\makecell{\#Params\\(M)}}\\
& & SSIM $\uparrow$ & PSNR $\uparrow$ & LPIPS $\downarrow$ & Ours & Fast & UltraFast & Ours & Fast & UltraFast & Ours & Fast & \\
\midrule

\multirow{3}{*}{2e-4}
& train & 0.7564 & 21.07 & 0.3102 & 0.97 & 1.04 & 1.55 & 368 & 362 & 200 & 1.92 & 0.59 & 0.06 \\
& truck & 0.8372 & 24.33 & 0.2376 & 1.01 & 1.10 & 1.59 & 351 & 345 & 190 & 1.87 & 0.57 & 0.06 \\
& \textbf{AVG} & 0.7968 & 22.70 & 0.2739 & 0.99 & 1.07 & 1.57 & 359 & 353 & 195 & 1.89 & 0.58 & 0.06 \\
\midrule

\multirow{3}{*}{1e-4}
& train & 0.7762 & 21.29 & 0.2823 & 1.46 & 1.61 & 2.54 & 407 & 370 & 196 & 2.02 & 0.79 & 0.09 \\
& truck & 0.8555 & 24.69 & 0.2039 & 1.53 & 1.70 & 2.57 & 389 & 371 & 194 & 1.98 & 0.71 & 0.10 \\
& \textbf{AVG} & 0.8158 & 22.99 & 0.2431 & 1.49 & 1.65 & 2.56 & 398 & 370 & 195 & 2.00 & 0.75 & 0.10 \\
\midrule

\multirow{3}{*}{5e-5}
& train & 0.7886 & 21.58 & 0.2641 & 2.01 & 2.21 & 3.53 & 379 & 385 & 203 & 1.95 & 0.87 & 0.13 \\
& truck & 0.8663 & 24.94 & 0.1798 & 2.23 & 2.45 & 3.97 & 380 & 381 & 200 & 2.51 & 0.84 & 0.15 \\
& \textbf{AVG} & 0.8274 & 23.26 & 0.2220 & 2.12 & 2.33 & 3.75 & 379 & 383 & 201 & 2.23 & 0.85 & 0.14 \\
\midrule

\multirow{3}{*}{2.5e-5}
& train & 0.7968 & 21.63 & 0.2490 & 2.81 & 2.92 & 4.73 & 423 & 390 & 205 & 3.02 & 1.37 & 0.17 \\
& truck & 0.8721 & 25.16 & 0.1679 & 2.99 & 3.29 & 5.29 & 414 & 398 & 208 & 2.74 & 1.26 & 0.19 \\
& \textbf{AVG} & 0.8344 & 23.39 & 0.2084 & 2.90 & 3.11 & 5.01 & 418 & 394 & 207 & 2.88 & 1.32 & 0.18 \\

\bottomrule
\end{tabular}
}
\end{table*}

\begin{table*}[t]
\centering
\scriptsize
\setlength{\tabcolsep}{4pt}
\renewcommand{\arraystretch}{1.15}
\caption{Per-scene results of SpeedyGS on the DeepBlending dataset. Note that the decoding time of the UltraFast version is nearly zero.}
\label{tab:each_scene_results_deepblending}
\resizebox{\textwidth}{!}{
\begin{tabular}{c|c|ccc|ccc|ccc|cc|cc}
\toprule
\multirow{2}{*}{$\lambda$} & \multirow{2}{*}{Dataset} & \multicolumn{3}{c|}{Fidelity} & \multicolumn{3}{c|}{Size $\downarrow$} & \multicolumn{3}{c|}{Training Time $\downarrow$} & \multicolumn{2}{c|}{Decoding Time $\downarrow$} & \multirow{2}{*}{\makecell{\#Params\\(M)}} \\
& & SSIM $\uparrow$ & PSNR $\uparrow$ & LPIPS $\downarrow$ & Ours & Fast & UltraFast & Ours & Fast & UltraFast & Ours & Fast & \\
\midrule

\multirow{3}{*}{2e-4}
& drjohnson    & 0.8821 & 28.65 & 0.3143 & 0.81 & 0.86 & 1.06 & 365 & 370 & 216 & 1.87 & 0.57 & 0.04\\
& playroom     & 0.8905 & 29.36 & 0.3129 & 0.59 & 0.63 & 0.66 & 334 & 331 & 190 & 1.78 & 0.44 & 0.03 \\
& \textbf{AVG} & 0.8863 & 29.01 & 0.3136 & 0.70 & 0.74 & 0.86 & 350 & 351 & 203 & 1.83 & 0.50 & 0.03\\
\midrule

\multirow{3}{*}{1e-4}
& drjohnson    & 0.8927 & 29.02 & 0.2923 & 1.21 & 1.30 & 1.87 & 427 & 382 & 216 & 2.07 & 0.61 & 0.07\\
& playroom     & 0.9007 & 30.06 & 0.2909 & 0.83 & 0.90 & 1.15 & 382 & 341 & 191 & 1.92 & 0.52 & 0.04\\
& \textbf{AVG} & 0.8967 & 29.54 & 0.2916 & 1.02 & 1.10 & 1.51 & 405 & 362 & 203 & 2.00 & 0.57 & 0.06\\
\midrule

\multirow{3}{*}{5e-5}
& drjohnson    & 0.8992 & 29.53 & 0.2777 & 1.75 & 1.98 & 2.98 & 407 & 405 & 223 & 2.01 & 0.70 & 0.10\\
& playroom     & 0.9065 & 30.25 & 0.2767 & 1.19 & 1.28 & 1.83 & 370 & 361 & 195 & 1.92 & 0.56 & 0.07\\
& \textbf{AVG} & 0.9028 & 29.89 & 0.2772 & 1.47 & 1.63 & 2.40 & 388 & 383 & 209 & 1.97 & 0.63 & 0.08\\
\midrule

\multirow{3}{*}{2.5e-5}
& drjohnson    & 0.9036 & 29.66 & 0.2663 & 2.54 & 2.84 & 4.43 & 431 & 426 & 231 & 2.27 & 0.86 & 0.15\\
& playroom     & 0.9097 & 30.48 & 0.2665 & 1.64 & 1.80 & 2.66 & 393 & 384 & 202 & 2.03 & 0.70 & 0.09\\
& \textbf{AVG} & 0.9066 & 30.07 & 0.2664 & 2.09 & 2.32 & 3.55 & 412 & 405 & 217 & 2.15 & 0.78 & 0.12\\

\bottomrule
\end{tabular}
}
\end{table*}

\end{document}